\documentclass[prd,aps,showpacs,preprintnumbers,amssymb]{revtex4}
\usepackage{graphicx}% Include figure files
\usepackage{MnSymbol}
\usepackage{epsf}
\def\e3p{$\eta \rightarrow 3 \pi$}

\begin{document}

\title{%
\hfill{\normalsize\vbox{%
\hbox{}
 }}\\
{
Proximity of $f_0(1500)$ and $f_0(1710)$ to the scalar glueball}}

\author{Amir H.~Fariborz,
$^{\it \bf a}$~\footnote[1]{Email:
 fariboa@sunyit.edu}}

\author{Azizollah Azizi
$^{\it \bf b}$~\footnote[5]{Email:
   azizi@shirazu.ac.ir }}

\author{Abdorreza Asrar
$^{\it \bf b}$~\footnote[2]{Email:
   ar\_asrar@shirazu.ac.ir   }}

\affiliation{$^ {\bf \it a}$ Department of
Mathematics and Physics,
State University of New York, Polytechnic Institute, Utica, NY 13502, USA.}

\affiliation{$^ {\bf \it b}$ Department of Physics,  Shiraz University, Shiraz, Iran, }

\date{\today}

\begin{abstract}
Within a nonlinear chiral Lagrangian framework,    the underlying mixings among quark-antiquark, four-quark and  glue components of $f_0(1500)$ and $f_0(1710)$ are studied in a global picture that includes all isosinglet scalar mesons below 2 GeV.  The quark components are introduced in the Lagrangian in terms of two separate nonets (a quark-antiquark nonet and a four-quark nonet) which can mix with each other and with a scalar glueball.
An iterative Monte Carlo simulation is developed to study the 14 free parameters of the Lagrangian by a simultaneous fit to more than 20 experimental data and constraints on the mass spectrum, decay widths, and decay ratios  of the isosinglet scalars below 2 GeV. Moreover, constraints on the mass spectrum and decay widths of isodoublet and isovector scalars below 2 GeV as well as pion-pion scattering amplitude are also taken into account.  In the leading order of the model
and within the overall experimental uncertainties, the ranges of variation of the model parameters are determined.   This leads to a set of points in the 14-dimensional parameter space at which the overall disagreement with experiment is no larger than the overall experimental uncertainties.    The insights gained in this global picture, due to the complexities of the mixings as well as the experimental uncertainties,   are mainly qualitative but are relatively robust,  and reveal that the lowest scalar glueball hides between $f_0(1500)$ and $f_0(1710)$, resulting in a considerable mixing with various quark components of these two states.    The overall current experimental and theoretical uncertainties do not allow to pin  down the exact glue components of isosinglet states,  nevertheless it is shown that the  $f_0(1500)$  and $f_0(1710)$ have the highest glue component.
While this global study does not allow precision predictions for each individual state,  it provides useful ``family''  correlations among the isosinglet states that are found insightful in probing the substructure of scalars, in general,  and the isosinglets,  in particular.    Specifically,  a close correlation between the substructure of isosinglets below and above 1 GeV is observed.  It is shown that as the simulations approach  the limit where  the  $f_0(500)$ and $f_0(980)$ become the two isosinglet  members of an ideally mixed two-quark two-antiquark nonet (which is widely believed to be a good approximation), the $f_0(1500)$ develops a large glue component.   The overall estimate of the scalar glueball mass is found to be $1.58 \pm 0.18$ GeV.

\end{abstract}

\pacs{14.80.Bn, 11.30.Rd, 12.39.Fe}

\maketitle
\section{introduction}

Developing a complete theory for low-energy QCD has been a great challenge for the theoretical particle physics and despite all the progress made over the past several decades the issue continues to linger in the field \cite{pdg}.   Among many challenges,   the fascinating phenomenon  of bound states formed out of gluons (glueballs) awaits both experimental verification as well as a model independent theoretical prediction.     In the meantime,  the quest for identifying which of the existing isosinglet scalar mesons are likely to have mixing with scalar glueball(s) continues.  But this identification is quite nontrivial since it gets  closely tied up with the issue of QCD vacuum and the breakdown of chiral symmetry via formation of quark-antiquarks,  four-quarks and gluonic condensates, and as a result, has rendered the achievement of this objective rather beyond the immediate reach of  the present state of the knowledge of low-energy QCD.   Despite all the existing complications,  the phenomenology of the mixing patterns among quark-antiquarks, four-quarks and glue components provides an insightful window into the world of scalar mesons in general,  and  the gluonic bound states, in particular.

Lattice QCD \cite{L1}-\cite{SVW_1995} provides a fundamental approach to understanding non-perturbative  QCD.    In the quenched approximation,  the glueball spectrum has been investigated in \cite{Morningstar_99} in which it is found that the lowest lying scalar glueball has a mass of 1730 (50)(80) MeV.   It is then important to know whether the unquenching effects are small or large.     Also, in order to verify whether the lattice predictions for glueballs actually coincide with observed hadrons, the mixing of pure glue with quark states will inevitably come into play,  and that naturally makes the study more complicated.   Nevertheless, a great amount of progress has been made on scalar mesons in lattice QCD
\cite{Wagner:2013nta}-\cite{Kunihiro:2003yj}.    Other fundamental approaches to study of glueballs include Bethe-Salpeter equation \cite{Bethe-Salpeter}, AdS and holography \cite{holography} and light-front approach \cite{light-front}.

To study the scalar glueball spectrum, various experimental data on scalar mesons, particularly those above 1 GeV are crucial.    In addition to the mass spectrum and decay properties given in PDG \cite{pdg},  several decay ratios are provided by the WA102 collaboration \cite{WA102}.    Moreover, recent data by LHCb \cite{LHCb1,LHCb2} on decays of  ${\bar B}^0$ and ${\bar B}_s^0$ to $J/\psi \pi^+ \pi^-$ has provided important probes of  the isosinglet states. In a recent report by BES III \cite{BESIII} the overlap of $f_0(1500)$ and $f_0(1710)$ with scalar glueball is studied in partial wave analysis of $J/\psi \rightarrow \gamma\eta\eta$.  It is also suggested \cite{Lu:2013jj,Minkowski:2004xf} that the production mechanism of heavy isosinglet states in such B decays are useful sources of information for identification of the scalar glueball.

Unlike the light pseudoscalars, understanding the properties of scalar mesons (particularly, their quark substructure) is known to be quite nontrivial and has become a serious puzzle for low-energy QCD.
Below 1 GeV, these states are \cite{pdg}:  The two isosinglets $f_0(500)$ [or sigma ($\sigma$) meson] and $f_0(980)$, the isodoublet $K_0^*(800)$ [or kappa ($\kappa$) meson] and the isotriplet $a_0(980)$.   It is known that a simple $q {\bar q}$ picture does not explain the properties of these mesons. The MIT bag model of Jaffe  \cite{Jaf} provides a foundation for understanding these states based on a four-quark
description and has inspired numerous investigations. Other models that investigate the nature of scalar mesons include $K {\bar K}$ molecule \cite{Isg}, unitarized quark model
\cite{vanBeveren1,vanBeveren2,Tor}, QCD sum-rules \cite{Eli}, chiral Lagrangians
\cite{JPGR_PRD84,RuizdeElvira:2010cs,04_ACCGL,Bardeen:2003qz,BFSS1,HSS1}
as well as many others \cite{Weinberg_13}-\cite{10_AOR}.
Above 1 GeV, the scalar states are \cite{pdg}:  the three isosinglet states $f_0(1370)$, $f_0(1500)$ and $f_0(1710)$, the isodoublet $K_0^*(1430)$ and the isotriplet $a_0(1450)$.   These states are generally believed to be closer to quark-antiquark states with two of them,  $f_0(1500)$ and $f_0(1710)$, strongly mix with glue.

An indispensable complication in understanding the physics of scalar glueballs is their mixing with isosinglet scalar mesons and that prevents their study to be ``standalone.''    This naturally brings into the discussion  the already messy situation of light scalar mesons within a region where no complete theoretical framework currently exists.    Fortunately, the established principles  of low-energy QCD (chiral symmetry and mechanisms of its breakdown, U(1)$_{\rm A}$ anomaly, large $N_c$ approximation, $\ldots$) have provided reliable guidelines  in exploring this region and have become the underlying platform upon which different low-energy QCD frameworks are built.   Chiral perturbation theory (ChPT) \cite{ChPT} has provided a systematic approach to the physics of strongly interacting pions and has led to many important developments.  However, the effectiveness of ChPT is near the threshold where the pion momenta are small,  and faces new challenges at higher energies where the strong influence of other resonances such as the rho meson, the sigma meson, $\ldots$ start to become the ``800 pound gorillas in the room.''    Attempts have been made to extend the domain of ChPT to include the effects of resonances \cite{Alarcon:2013taa}-\cite{Ecker:1988te}, which are manifested indirectly through low-energy constants at order ${\cal O} p^4$ \cite{Ecker:1988te}.

Since the objective of this work is to explore the properties of scalar mesons, it is natural to keep the scalar fields explicitly in the Lagrangian instead of integrating them out. We work within a nonlinear chiral Lagrangian framework  that includes scalar fields below and above 1 GeV.      In this approach,   various low-energy processes including various scattering and decays have been successfully studied   \cite{e3p}-\cite{06_F}.  The scattering amplitudes can be  reasonably well approximated by the contact terms describing the interaction of pseudodscalars together with tree level diagrams representing the contribution of intermediate resonances up to slightly above the energy region of interest.     It is observed  that even though the contribution of the individual diagrams may be large and violate the expected unitarity bounds, they balance each other in such a way that the unitarity bounds are respected.  The details also involve additional effects such as regularization of the pole diagrams by inclusion of imaginary parts (determined by fits to experimental data) which can be interpreted as the decay width of the resonances and hence are subleading in $1/N_c$ expansion.  Another subleading effect is inclusion of four-quark scalar mesons which is one of the focuses of the present work and is described in next section in more detail.

There is considerable evidence that the scalars  below 1 GeV are not simple quark-antiquark states and perhaps are closer to a four-quark picture (either of molecular types or of tetraquark types or a combination of the two) whereas those above 1 GeV are closer to the conventional $p$-wave quark-antiquark states.  At the same time, more refined analysis shows that the situation with such
four-quark versus quark-antiquark is not that clear cut and both sets of states have some distortions from those pure pictures.   It is therefore natural to ask whether such distortions are generated by an underlying  mixing among scalar states below and above 1 GeV.   There seems to be a clear evidence for a mixing scenario as was pointed out in \cite{Mec} for the isodoublets $K_0^*(800)$ and $K_0^*(1430)$,  as well as isotriplets $a_0(980)$ and $a_0(1450)$ and further extended to the more complicated cases of isosinglets $f_0(500)$, $f_0(980)$, $f_0(1370)$, $f_0(1500)$ and $f_0(1710)$  in Refs.\ \cite{04_F1,06_F}.   The complications of dealing with isosinglet scalar states are due to their underlying mixings; the two isosinglet states within each scalar meson nonet  not only can mix with each other as well as  with the isosinglets of the other nonet,  they also can, in addition,  mix with one (or more) scalar glueball(s).   This leads, at the very least, to a 5$\times$5 mixing as opposed to the 2$\times$2 cases for isodoublets and isotriplets.     Other investigators have studied different  aspects of scalar meson mixings \cite{NR04,Teshima02,close02}.

Finally,  we point out that the investigations of the scalar mesons in the nonlinear chiral Lagrangian framework of \cite{e3p}-\cite{06_F}, which are the basis of this study,  have resulted in an overall  coherent picture for the light scalar mesons.  Particularly the  substructure of the scalar mesons below and above 1 GeV have shown close correlations which can be best understood by probing their underlying mixings.  It is found that the scalars below 1 GeV being closer to four-quark states and those above 1 GeV closer to quark-antiquark states.  This  has been a robust outcome of these studies. In addition,  the results for the underlying mixing patterns and substructures of scalar mesons obtained in the nonlinear chiral Lagrangians of \cite{e3p}-\cite{06_F} are in a close agreement with similar results found within the context of the linear sigma model \cite{3flavor}-\cite{LsM} which gives added confidence and further motivates the present work.

In this  article,  we extend the previous works on isosinglet mixing of ref.~\cite{06_F}  by performing a simultaneous fit to the mass and decay properties of isosinglet states.    As we shall see this will involve exploring the 14-dimensional parameter space of the model using an iterative Monte Carlo simulation of the available experimental data.  This  will result in determining the range of variation of the 14 model parameters, which will subsequently be used to explore the quark and glue substructure of isosinglet scalars.   In Sec.\ II, we give a brief review of the model followed by our strategy for determining the model parameters in Sec.\ III, and  the numerical results (for global simulation I) in Sec.~IV.   We then give a discussion of the sensitivity of the results on the experimental inputs in Sec.~V, followed by the properties of scalar glueball in VI.   We end by giving a summary and discussion of the results in Sec.\ VII.

%%%%%%%%%%%%%%%%%%%%%%%%%%%%%%%%%%%%%%%%%%%%%%%

\section{Brief Review of the model}

%%%%%%%%%%%%%%%%%%%%%%%%%%%%%%%%%%%%%%%%%%%%%%%
Since the lightest scalars [$f_0(500)$, $K_0^*(800)$, $f_0(980)$ and $a_0(980)$] do not form a pure quark-antiquark nonet, the  interesting question is whether the next-to-lying set of scalars [$f_0(1370)$, $K_0^*(1430)$, $a_0(1450)$, $f_0(1500)$, $f_0(1710)$] are the right candidates for such a nonet.    In fact, it is speculated that there is a quark-antiquark scalar meson nonet above 1 GeV with the $K_0^*(1430)$ and the $a_0(1450)$ its  likely members.   However, a close look at the properties of these two states raises some serious questions for this assignment.      For example,  in a $q {\bar q}$ nonet, the isotriplet is expected
to be lighter than the isodoublet, but for these two
states \cite{pdg}:
\begin{eqnarray}
m \left[ a_0(1450) \right] &=& 1474 \pm 19 \hskip .2cm \mathrm{MeV}, \nonumber \\
m \left[ K_0^*(1430) \right] &=& 1425 \pm 50  \hskip .2cm \mathrm{MeV},
\label{a0k0_mass_exp}
\end{eqnarray}
which are comparable at best.  Also their decay ratios given in PDG \cite{pdg} do not follow a pattern
expected from an SU(3) symmetry \cite{Mec} (given in parenthesis):
\begin{eqnarray}
{ {\Gamma \left[ a_0(1450) \right]}\over
{\Gamma\left[K_0^*(1430)\right] }} &=& 0.98 \pm 0.34
\hskip .2cm (1.51), \nonumber \\
\hskip 0.2cm
{ {\Gamma \left[ a_0(1450)\rightarrow K{\bar K} \right]}\over
{\Gamma\left[a_0(1450)\rightarrow\pi \eta\right] }} &=& 0.88 \pm 0.23
\hskip .2cm  (0.55),\nonumber \\
\hskip 0.2cm
{ {\Gamma \left[ a_0(1450)\rightarrow \pi\eta' \right]}\over
{\Gamma\left[a_0(1450)\rightarrow\pi \eta\right] }} &=& 0.35\pm0.16 \hskip
.2cm  (0.16).
\label{a0k0_decay_exp}
\end{eqnarray}
These deviations suggest that these states are likely to have small deviations from pure quark-antiquark states.
The next natural question is whether these deviations are the result of   underlying mixings with a four-quark nonet below 1 GeV.   This question was taken up in Ref.\ \cite{Mec} in which it was shown how a simple global picture for the
scalar mesons below and above 1 GeV can originate from a mixing between a four-quark nonet $N$  below 1 GeV and a quark-antiquark nonet $N'$ above 1 GeV.    It is shown in \cite{Mec} that allowing these two nonets to slightly mix, leads to a natural explanation for the deviations in mass [Eq.~(\ref{a0k0_mass_exp})] and decay properties [Eq.~(\ref{a0k0_decay_exp})].   The theoretical framework used in \cite{Mec} is a nonlinear chiral Lagrangian in which the mass terms for the isodoublet and isotriplet states can be written in terms of the two scalar meson nonets
\begin{equation}
{\cal L}_\mathrm{mass}^{I=1/2,1} =
- a\ \mathrm{Tr}(NN) - b\ \mathrm{Tr}(NN{\cal M})
- a'\ \mathrm{Tr}(N'N') - b'\ \mathrm{Tr}(N'N'{\cal M}),
\label{L_mass_I1}
\end{equation}
where ${\cal M} = \mathrm{diag} (1,1,x)$ with $x$ being the ratio of
the strange to non-strange quark masses, and $a,b,a'$ and $b'$ are
the free parameters determined  by the ``bare'' (unmixed)
masses of $I=1/2$ and $I=1$ scalars
\begin{equation}
\begin{array}{cclcccl}
m^2 [a_0] &=& 2 (a + b), & &  m^2 [a'_0] & = & 2 (a' + b'),\\
m^2 [K_0] &=& 2a + (1+x) b, & &  m^2 [K'_0] & = & 2 a' + (1+x) b'.
\end{array}
\label{Bare_mass_eq_I12}
\end{equation}
where the subscript ``0'' denotes the ``bare'' masses.
The model of \cite{Mec} assumes that the light four-quark nonet  $N$ is lower in mass than the quark-antiquark nonet $N'$.  Taking into account the fact that in a conventional quark-antiquark nonet the isodoublet (that has one strange quark) is lighter than the isotriplet (that does not have an strange quark), and together with the fact that in a four-quark nonet this spectrum is reversed (i.e. the isodoublet has one strange quark and therefore is lighter than the isotriplet with two strange quarks), we expect:
\begin{equation}
m^2 [K_0] <  m^2 [a_0] \le m^2 [a'_0] < m^2 [K'_0].
\label{mass_order}
\end{equation}
This mass ordering is our indirect connection to the quark content of $N$ and $N'$.  The leading  mixing of these two nonets is described by 
\begin{equation}
{\cal L}_\mathrm{mix}^{I=1/2,1} = -\gamma \mathrm{Tr} \left( N N' \right),
\label{L_mix_I1}
\end{equation}
where, with a detailed numerical analysis,   it is shown in \cite{Mec} that  for
$0.51 < \gamma < 0.62 \hskip .2cm\mathrm{GeV^2}$, it is possible to
describe the physical masses such that the bare masses uphold
the expected ordering of (\ref{mass_order}).
Specifically, using the fact that when two states with ``bare'' masses $M_I < M_{II}$ slightly mix, the resulting physical masses (${\tilde M}_I < {\tilde M}_{II}$) split away from the two ``bare'' masses (i.e. ${\tilde M}_I < M_I <M_{II} < {\tilde M}_{II}$), and that the amount of splitting is inversely proportional to the ``bare'' mass difference $M_{II}^2 - M_I^2$.   Hence, the two ``bare'' isotriplet states [which are closer to each other according to (\ref{mass_order}] split more
than the two isodoublet states, and
consequently, the
physical isovector state $a_0(1450)$ becomes heavier than the
isodoublet state $K_0^*(1430)$  in agreement with the observed
experimental values in
(\ref{a0k0_mass_exp}).  The
light isovector and
isodoublet states are the $a_0(980)$ and the
$K_0^*(800)$.    With the physical masses $m[a_0(980)]=0.9835$ GeV,
$m [K_0^*(800)] =0.875$ GeV, $m[a_0(1450)]=1.455$ GeV and
$m[K_0^*(1430)]=1.435$ GeV, the best values of mixing parameter $\gamma$ and
the bare
masses are found in \cite{Mec}
\begin{equation}
m\left[ a_0\right] =m\left[ a'_0\right] =1.24 \hskip .1cm\mathrm{GeV},\hskip .2cm
m\left[ K_0\right] =1.06 \hskip .1cm\mathrm{GeV}, \hskip .2cm
m\left[ K'_0\right] =1.31 \hskip .1cm\mathrm{GeV}, \hskip .2cm
\gamma = 0.58 \hskip .1cm \mathrm{GeV^2}.
\label{a0k0_m_bare}
\end{equation}
The decay properties of the isodoublet and isotriplet states are also studied in \cite{Mec}.    The Lagrangian for the coupling of the two nonets $N$ and $N'$ to two-pseudoscalar particles
are then studied  and its unknown parameters are found by comparing with available experimental data on relevant  decay widths.
The work of Ref.\ \cite{Mec} shows that the $I=1$ states are close
to maximal mixing [i.e.\ $a_0(980)$ and $a_0(1450)$ are
approximately made of 50\% quark-antiquark and 50\%
four-quark components], and the $I=1/2$ states have a
similar structure with $K_0^*(800)$ made of
approximately 74\% of four-quark and 26\%
quark-antiquark, and the reverse structure for the
$K_0^*(1430)$.

The interaction Lagrangian density for the $I=1/2,\ 1$ scalars is developed in \cite{Mec}
\begin{equation}
{\cal L}_{\rm int}^{I=1/2,1} =
 A \epsilon^{abc}\epsilon_{def}
N_a^d\partial_\mu\phi^e_b\partial_\mu\phi^f_c
+ C {\rm Tr} (N \partial_\mu \phi) {\rm Tr} (\partial_\mu\phi)
+ A' \epsilon^{abc}\epsilon_{def}
{N'}_a^d\partial_\mu\phi^e_b\partial_\mu\phi^f_c
+ C' {\rm Tr} (N' \partial_\mu \phi) {\rm Tr} (\partial_\mu\phi),
\label{L_int_I1}
\end{equation}
where  $A$, $C$, $A'$ and $C'$ are the a priori
unknown parameters fixed by experimental inputs on mass and decay properties of $I=1/2$ and $I=1$ states below and above 1 GeV.  It is found in  \cite{Mec} that
\begin{equation}
A=1.19\pm 0.16 \hskip .1cm {\rm GeV}^{-1}, \hskip 0.2cm
A'=-3.37\pm 0.16 \hskip .1cm  {\rm GeV}^{-1}, \hskip 0.2cm
C=1.05\pm 0.49 \hskip .1cm  {\rm GeV}^{-1}, \hskip 0.2cm
C'=-6.87\pm 0.50 \hskip .1cm  {\rm GeV}^{-1}. \hskip 0.2cm
\label{AC_values}	
\end{equation}

The case of $I=0$ states in this mixing mechanism is considerably more
complicated due to their mixing with scalar glueball(s).
The mixing model of Ref.\ \cite{Mec} was further extended to include $I=0$ states in
Refs.\ \cite{04_F1,04_F2,06_F}.   The general mass terms for $I=0$ states and a scalar glueball $G$
can be written as:
\begin{eqnarray}
{\cal L}_\mathrm{mass}^{I=0}  &=& {\cal L}_\mathrm{mass}^{I=1/2,1}
- c\ \mathrm{Tr}(N)\mathrm{Tr}(N)
- d\ \mathrm{Tr}(N) \mathrm{Tr}(N{\cal M})
\nonumber \\
&&- c'\ \mathrm{Tr}(N')\mathrm{Tr}(N')
- d'\ \mathrm{Tr}(N') \mathrm{Tr}(N'{\cal M})
- {1\over 2}m_G^2\, G^2.
\label{L_mass_I0}
\end{eqnarray}
It is easy to see that the role of terms involving $c$ and $d$ parameters is to induce ``internal'' mixing
between the two $I=0$ flavor combinations [$(N_1^1 + N_2^2)/\sqrt{2}$ and
$N_3^3$] of nonet $N$ (a similar role is played by terms involving  $c'$ and $d'$ in nonet $N'$).
Parameters $c,d,c'$ and $d'$ do not contribute to
the mass spectrum of the $I=1/2$ and $I=1$ states.
The last term represents the glueball mass term (justifications for identifying field $G$ with an scalar glueball is discussed in detail in Sec.~VI). It is seen that part of the mass Lagrangian for isosinglet states is constrained by the mass term for isodoublets and isotriplets, i.e. the term ${\cal L}_\mathrm{mass}^{I=1/2,1}$ discussed in Eq. ~(\ref{L_mass_I1}) with its parameters determined in Eqs. ~(\ref{Bare_mass_eq_I12}) and (\ref{a0k0_m_bare}).

The mixing between $N$ and
$N'$, and the mixing of these two nonets with the  scalar glueball $G$ can
be
written as
\begin{equation}
{\cal L}_\mathrm{mix}^{I=0} =
{\cal L}_\mathrm{mix}^{I=1/2,1}
- \rho\ \mathrm{Tr} (N) \mathrm{Tr} (N')
- e G\ \mathrm{Tr} \left( N \right)
- f G\ \mathrm{Tr} \left( N' \right),
\label{L_mix_I0}
\end{equation}
where the first term is given in
(\ref{L_mix_I1}) with
$\gamma$ from (\ref{a0k0_m_bare}).   The second term on the right hand side does not
contribute to the $I=1/2,\ 1$ mixing, and terms  with unknown couplings $e$ and $f$
describe mixing with the scalar glueball $G$ (also not contributing to $I=1/2,\ 1$ cases).
As a result, the five isosinglets
below
2 GeV, become a mixture of five different flavor combinations, and their
masses can be organized as
\begin{equation}
{\cal L}_\mathrm{mass}^{I=0} + {\cal L}_\mathrm{mix}^{I=0} = - {1\over 2} {\tilde {\bf
F}}_0 {\bf M}^2 {\bf F}_0 =
- {1\over 2} {\tilde {\bf F}} {\bf M}_\mathrm{diag}^2 {\bf F},
\label{L_mass_mix}
\end{equation}
with
\begin{equation}
{\bf F}_0 =
       \begin{array}{l}
                         \left(
                         \begin{array}{c}
                              N_3^3 \\
                             (N_1^1 + N_2^2)/\sqrt{2}\\
                              {N'}_3^3 \\
                             ({N'}_1^1 + {N'}_2^2)/\sqrt{2}\\
                              G
                         \end{array}
                         \right)

       \end{array}
=
       \begin{array}{l}
                         \left(
                         \begin{array}{l}
                             f_0^\mathrm{NS}\\
                             f_0^\mathrm{S} \\
                             {f'}_0^\mathrm{S}\\
                             {f'}_0^\mathrm{NS} \\
                              G
                         \end{array}
                         \right)
       \end{array}
\propto
       \begin{array}{l}
                         \left(
                         \begin{array}{c}
      {\bar u}{\bar d} u d\\
     ({\bar d}{\bar s} d s + {\bar s}{\bar u} s u) /\sqrt{2}\\
                              s{\bar s}\\
                             (u{\bar u}  + d {\bar d})/\sqrt{2}\\
                             \alpha_s G^{\mu\nu}G_{\mu\nu}
                         \end{array}
                         \right),

       \end{array}
\label{SNS_def}
\end{equation}
where the superscript NS and S respectively represent the non-strange
and strange combinations.  ${\bf F}$ contains the physical fields
\begin{equation}
{\bf F}  =
       \begin{array}{c}
                        \left( \begin{array}{c}
                                 f_0(500)\\
                                f_0(980)\\
                                f_0(1370)\\
                                f_0(1500)\\
                                f_0(1710)
                                \end{array}
                         \right)
= K^{-1} {\bf F}_0,
       \end{array}
\label{K_def}
\end{equation}
where $K^{-1}$ is the transformation matrix.
The mass squared matrix is
\begin{equation}
{\bf M}^2  =
\left[
\begin{array}{ccccc}
%---------------------------------------------------------
2 m^2\left[K_0\right] - m^2\left[ a_0\right]  + 2 (c + d x) &
\sqrt{2}[2c + (1+x)d] &
\gamma + \rho &
\sqrt{2} \rho &
e
\\
%---------------------------------------------------------
\sqrt{2}[2c + (1+x)d] &
m^2\left[a_0\right]  + 4 (c + d) &
\sqrt{2} \rho &
\gamma + 2 \rho&
\sqrt{2} e
\\
%---------------------------------------------------------
\gamma + \rho&
\sqrt{2} \rho &
2 m^2\left[ K'_0\right] - m^2\left[ a'_0\right]  + 2 (c' + d' x) &
\sqrt{2}[2c' + (1+x)d'] &
f
%---------------------------------------------------------
\\
\sqrt{2} \rho &
\gamma + 2 \rho&
\sqrt{2}[2c' + (1+x)d'] &
m^2\left[ a'_0\right] + 4 (c' + d') &
\sqrt{2} f
\\
%---------------------------------------------------------
e &
\sqrt{2} e &
f &
\sqrt{2} f &
m_G^2
%---------------------------------------------------------
\end{array}
\right],
\label{mass_matrix}
\end{equation}
in which the value of the unmixed $I=1/2,\ 1$ masses, and
the mixing parameter $\gamma$ are substituted in from
(\ref{a0k0_m_bare}).   There are eight unknown parameters in
(\ref{mass_matrix}); $c$, $d$, $c'$, $d'$, $m_G$, $\rho$, $e$ and
$f$.

The scalar-pseudoscalar-pseudoscalar interaction takes the general form:
\begin{eqnarray}
{\cal L}_\mathrm{int}^{I=0} &=&
{\cal L}_\mathrm{int}^{I=1/2,1} +
B\ \mathrm{Tr} \left( N \right) \mathrm{Tr} \left({\partial_\mu}\phi
{\partial_\mu}\phi \right)
+ D\ \mathrm{Tr} \left( N \right) \mathrm{Tr}
\left({\partial_\mu}\phi \right)  \mathrm{Tr} \left( {\partial_\mu}\phi
\right)
+ B'\ \mathrm{Tr} \left( N' \right) \mathrm{Tr} \left({\partial_\mu}\phi
{\partial_\mu}\phi \right)
\nonumber\\
&& + D'\ \mathrm{Tr} \left( N' \right) \mathrm{Tr}
\left({\partial_\mu}\phi \right)  \mathrm{Tr} \left( {\partial_\mu}\phi
\right)
+ E G\  \mathrm{Tr} \left({\partial_\mu}\phi
{\partial_\mu}\phi \right)
+ F G\  \mathrm{Tr}
\left({\partial_\mu}\phi \right)  \mathrm{Tr} \left( {\partial_\mu}\phi
\right),
\label{L_int_I0}
\end{eqnarray}
where $B$ and $D$ are unknown coupling constants describing the
coupling of the four-quark nonet $N$ to the pseudoscalars.  Similarly,
$B'$ and $D'$  are couplings of $N'$ to the pseudoscalars.
$E$ and $F$
describe the coupling of the scalar glueball to the pseudoscalar mesons. The term
${\cal L}_\mathrm{int}^{I=1/2,1}$ is the interaction Lagrangian for $I=1/2,\ 1$
states given in Ref.\ \cite{BFSS1}.
For decay width computation we recast the interaction Lagrangian (\ref{L_int_I0}) in terms of physical fields and physical couplings
\begin{eqnarray}
- {\cal L}_\mathrm{int} &=&
{1\over \sqrt{2}} \gamma_{\pi\pi}^i \, {\bf F}_i \partial_\mu \pi \cdot
\partial_\mu
\pi
+ {1\over \sqrt{2}} \gamma_{KK}^i \, {\bf F}_i \partial_\mu {\bar K}
\partial_\mu K
\nonumber \\
&& + \gamma_{\eta\eta}^i \, {\bf F}_i \partial_\mu \eta
\partial_\mu \eta
+ \gamma_{\eta\eta'}^i \, {\bf F}_i \partial_\mu \eta
\partial_\mu \eta'
+ \gamma_{\eta'\eta'}^i \, {\bf F}_i \partial_\mu \eta'
\partial_\mu \eta',
\label{L_int_I0_b}
\end{eqnarray}
where $\gamma^i_{ss'}$ is the physical coupling of the $i$-th scalar [$i=1\ldots 5$; see Eq.~(\ref{K_def})] to
pseudoscalars $s$ and $s'$ given by
\begin{equation}
\gamma^i_{ss'} = \sum_j \left( \gamma_{ss'} K\right)_{ji},
\end{equation}
with  $K$ defined in  ({\ref{K_def}) and
$
\gamma_{ss'} = \mathrm{diag}
\left( \gamma^\mathrm{NS}_{ss'}, \gamma^\mathrm{S}_{ss'}, {\gamma'}^\mathrm{S}_{ss'}, {\gamma'}^\mathrm{NS}_{ss'},\gamma^G_{ss'} \right)
$
in which the diagonal elements are the couplings of the
pseudoscalars $s$ and $s'$ to the $f_0^\mathrm{NS}, f_0^\mathrm{S}, {f'}_0^\mathrm{S}, {f'}_0^\mathrm{NS}$
[defined in (\ref{SNS_def})] and the scalar glueball $G$, respectively.
The diagonal elements for all decay channels $ss'$ are listed in Ref.\ \cite{04_F1}.

%%%%%%%%%%%%%%%%%%%%%%%%%%%%%%%%%%%%%%%%%%%%%%%%%%%%%%

\section{Strategy for determining the model parameters}

%%%%%%%%%%%%%%%%%%%%%%%%%%%%%%%%%%%%%%%%%%%%%%%%%%%%%%
There are 14 unknown parameters in the $I=0$ part of the Lagrangian density that we need to determine by incorporating appropriate experimental data on the mass spectrum as well as the appropriate decay widths and decay ratios.      These can be divided into a six-dimensional parameter space ($B$, $D$, $B'$, $D'$, $E$ and $F$) that only affect the scalar-pseudoscalar-pseudoscalar coupling constants, and an eight-dimensional parameter space ($c$, $d$, $c'$, $d'$, $m_G$, $\rho$, $e$ and $f$) that both directly enter into the $5\times5$ mass matrix, and also indirectly enter in the calculation of decay widths and decay ratios through the rotation matrix $K$ that rotates the bare bases into the physical bases.    As a result,  determining these two groups of parameters independently of each other is only an approximation;  a general approach requires a simultaneous 14-parameter fit.    In works of Refs.\ \cite{04_F1,04_F2,06_F}, as a preliminary study,   these two parameter spaces were studied in separate fits in some details.  Here we generalize those findings by performing simultaneous fits using an iterative Monte Carlo algorithm developed by the authors \cite{FAA_paper3}.    This general method of dealing with this parameter space is considerably more complicated than those presented in  \cite{04_F1,04_F2,06_F}.    An added complication in this analysis is the lack of established experimental data  on some of the properties of scalar mesons.    To minimize the effect of unestablished experimental data,   different sets of target experimental quantities are considered.    The target quantities displayed in Table \ref{quantities1} are in three groups of masses,  partial decay widths and several decay ratios reported by WA102 collaboration \cite{WA102}; altogether forming 23 target quantities.    Some of these quantities are not firmly established:  The partial decay widths of $f_0(1370)$ to $\pi\pi$ and $K{\bar K}$ are reported in \cite{Bugg:2007ja} but  are not used in any averaging by PDG.   Also, in the analysis of WA102 collaboration \cite{WA102} the mass of $f_0(1370)$, $f_0(1500)$ and $f_0(1710)$ are fixed to the values displayed in Table \ref{quantities1} and are not the same as those given in PDG \cite{pdg} or in Refs. \cite{Bugg:2007ja,oller} (for example, the mass of $f_0(1370)$ given by PDG is in the broad range of 1.2--1.5 GeV, to be compared with fixed value of 1.312 GeV selected by WA102 collaboration or 1.300 $\pm$ 0.015 GeV reported in Ref.~\cite{Bugg:2007ja}). In our analysis we have been mindful of these variations and have minimized the uncertainties of our predictions that stem from such inputs,  by  considering three different simulations, each of which emphasizes a particular version of such inputs.  In our global simulation I,  we target the inputs of Table  \ref{quantities1} with the exception of the two partial decay widths of  $f_0(1370)$ (i.e.~in global simulation I, we target 21 out of the 23 quantities given in that table).   In our global simulation II,   we include all the 23 quantities displayed in Table \ref{quantities1} (in this case since the WA102 data forms a significant part of the experimental inputs, we use the mass of $f_0(1370)$ of WA102 collaboration which is not too different from that of \cite{Bugg:2007ja}).    In global simulation III,  we use the target quantities displayed in Table \ref{quantities2} in which we have   excluded the decay ratios of WA102 collaboration \cite{WA102} and in addition the mass of $f_0(1370)$, $f_0(1500)$ and $f_0(1710)$ are respectively taken from \cite{Bugg:2007ja}, PDG \cite{pdg} and \cite{oller}.

\begin{table}[t]
	\begin{center}
		\begin{tabular}{|c|c|c|}
			\hline
			Short notation & Quantity  &  Target value [Ref.] \\ \hline
			$m_1$  & $m[f_0(500)]$& 400--550  {\rm MeV} \cite{pdg} \\ \hline
			$m_2$ & $m[f_0(980)]$ & $990 \pm 20$ {\rm MeV} \cite{pdg}\\ \hline
			$m_3$ & $m[f_0(1370)]$ & $1312$  {\rm MeV} \cite{WA102} \\ \hline
			$m_4$ & $m[f_0(1500)]$&$ 1502$  {\rm MeV} \cite{WA102} \\ \hline
			$m_5$ & $m[f_0(1710)]$& $1727$  {\rm MeV} \cite{WA102} \\ \hline \hline
			$\Gamma^3_{\frac{\pi\pi}{KK}}$&$\frac{\Gamma[f_0(1370)\rightarrow\pi\pi]}{\Gamma[f_0(1370)\rightarrow K\bar{K} ]}$&$2.17  \pm  0.9$ \cite{WA102}\\ \hline
			 $\Gamma^3_{\frac{\eta\eta}{KK}}$&$\frac{\Gamma[f_0(1370)\rightarrow\eta\eta]}{\Gamma[f_0(1370)\rightarrow K\bar{K} ]}$&$0.35  \pm  0.30$  \cite{WA102}\\ \hline
			 $\Gamma^4_{\frac{\pi\pi}{\eta\eta}}$&$\frac{\Gamma[f_0(1500)\rightarrow\pi\pi]}{\Gamma[f_0(1500)\rightarrow \eta\eta ]}$& $5.56  \pm 0.93$ \cite{WA102}\\ \hline
			$\Gamma^4_{\frac{KK}{\pi\pi}}$&$\frac{\Gamma[f_0(1500)\rightarrow K\bar{K}]}{\Gamma[f_0(1500)\rightarrow \pi\pi ]}$& $0.33  \pm  0.07$  \cite{WA102}\\ \hline
			$\Gamma^4_{\frac{\eta \eta'}{\eta \eta}}$&$\frac{\Gamma[f_0(1500)\rightarrow\eta \eta' ]}{\Gamma[f_0(1500)\rightarrow \eta\eta ]}$&  $0.53  \pm  0.23$ \cite{WA102}\\ \hline
			$\Gamma^5_{\frac{\pi\pi}{KK}}$&$\frac{\Gamma[f_0(1710)\rightarrow\pi\pi]}{\Gamma[f_0(1710)\rightarrow K\bar{K} ]}$& $0.20  \pm  0.03$  \cite{WA102}\\ \hline
			$\Gamma^5_{\frac{\eta \eta}{KK}}$&$\frac{\Gamma[f_0(1710)\rightarrow\eta\eta]}{\Gamma[f_0(1710)\rightarrow K\bar{K} ]}$&$ 0.48  \pm   0.19$ \cite{WA102}\\ \hline \hline
			$\Gamma^1_{\pi\pi}$&$\Gamma[f_0(500)\rightarrow\pi\pi]$& 400--700 {\rm MeV} \cite{pdg}  \\ \hline
			$\Gamma^2_{\pi\pi}$&$\Gamma[f_0(980)\rightarrow\pi\pi]$& 40--100 {\rm MeV} \cite{pdg}\\ \hline
			$\Gamma^3_{\pi\pi}$&$\Gamma[f_0(1370)\rightarrow\pi\pi]$&$(0.26 \pm 0.09)\times(230 \pm 15)$ {\rm MeV} \cite{Bugg:2007ja}\\ \hline
			$\Gamma^3_{KK}$&$\Gamma[f_0(1370)\rightarrow K\bar{K}]$&$(0.35 \pm 0.13)\times(230 \pm 15)$ {\rm MeV} \cite{Bugg:2007ja}\\ \hline
			$\Gamma^4_{\pi\pi}$&$\Gamma[f_0(1500)\rightarrow\pi\pi]$& $(0.349 \pm 0.023)\times(109 \pm 7)$ {\rm MeV} \cite{pdg}\\ \hline
			$\Gamma^4_{KK}$&$\Gamma[f_0(1500)\rightarrow K\bar{K}]$& $(0.086 \pm 0.010 )\times(109 \pm 7)$ {\rm MeV} \cite{pdg} \\ \hline
			$\Gamma^4_{\eta\eta}$&$\Gamma[f_0(1500)\rightarrow\eta\eta]$& $(0.051 \pm 0.009)\times(109 \pm 7)$ {\rm MeV}  \cite{pdg}\\ \hline
			$\Gamma^4_{\eta\eta\prime}$&$\Gamma[f_0(1500)\rightarrow\eta\eta\prime]$& $(0.019 \pm 0.008)\times(109 \pm 7)${\rm MeV}\cite{pdg}\\ \hline
			$\Gamma^5_{\pi\pi}$&$\Gamma[f_0(1710)\rightarrow\pi\pi]$& $(0.12 \pm 0.11)\times(220 \pm 40)$ {\rm MeV} \cite{oller}\\ \hline
			$\Gamma^5_{KK}$&$\Gamma[f_0(1710)\rightarrow K\bar{K}]$& $(0.36\pm 0.12)\times(220 \pm 40)$ {\rm MeV} \cite{oller}\\ \hline
			$\Gamma^5_{\eta\eta}$&$\Gamma[f_0(1710)\rightarrow \eta\eta]$& $(0.22 \pm 0.12)\times(220 \pm 40)$ {\rm MeV} \cite{oller} \\ \hline
		\end{tabular}
	\end{center}
	\caption{Target quantities used to explore the 14 parameters of the Lagrangian in global simulations I and II.  In global  simulation I (II) the decay channels of $f_0(1370)$  are excluded (included).  The short notation for the quantities are defined in column one.}
	\label{quantities1}
\end{table}

% % % % % % % % % % % % % % % % % % % % % % % % % %
\begin{table}[t]
	\begin{center}
		\begin{tabular}{|c|c|c|}
			\hline
			Short notation & Quantity &  Target  value [Ref.] \\ \hline
			$m_1$  & $m[f_0(500)]$& 400--550  {\rm MeV} \cite{pdg} \\ \hline
			$m_2$ & $m[f_0(980)]$ & $990 \pm 20$ {\rm MeV} \cite{pdg}\\ \hline
			$m_3$ & $m[f_0(1370)]$ & $1300 \pm 15$  {\rm MeV} \cite{Bugg:2007ja} \\ \hline
			$m_4$ & $m[f_0(1500)]$&$ 1505 \pm 6$  {\rm MeV} \cite{pdg} \\ \hline
			$m_5$ & $m[f_0(1710)]$& $1690 \pm 20$  {\rm MeV} \cite{oller} \\ \hline \hline
			$\Gamma^1_{\pi\pi}$&$\Gamma[f_0(500)\rightarrow\pi\pi]$& 400--700 {\rm MeV} \cite{pdg}  \\ \hline
			$\Gamma^2_{\pi\pi}$&$\Gamma[f_0(980)\rightarrow\pi\pi]$&  40--100 {\rm MeV} \cite{pdg}\\ \hline
			$\Gamma^3_{\pi\pi}$&$\Gamma[f_0(1370)\rightarrow\pi\pi]$&$(0.26 \pm 0.09)\times(230 \pm 15)$ {\rm MeV} \cite{Bugg:2007ja}\\ \hline
			$\Gamma^3_{KK}$&$\Gamma[f_0(1370)\rightarrow K\bar{K}]$&$(0.35 \pm 0.13)\times(230 \pm 15)$ {\rm MeV} \cite{Bugg:2007ja}\\ \hline
			$\Gamma^4_{\pi\pi}$&$\Gamma[f_0(1500)\rightarrow\pi\pi]$& $(0.349 \pm 0.023)\times(109 \pm 7)$ {\rm MeV} \cite{pdg}\\ \hline
			$\Gamma^4_{KK}$&$\Gamma[f_0(1500)\rightarrow K\bar{K}]$& $(0.086 \pm 0.010 )\times(109 \pm 7)$ {\rm MeV} \cite{pdg} \\ \hline
			$\Gamma^4_{\eta\eta}$&$\Gamma[f_0(1500)\rightarrow\eta\eta]$& $(0.051 \pm 0.009)\times(109 \pm 7)$ {\rm MeV}  \cite{pdg}\\ \hline
			$\Gamma^4_{\eta\eta\prime}$&$\Gamma[f_0(1500)\rightarrow\eta\eta\prime]$& $(0.019 \pm 0.008)\times(109 \pm 7)${\rm MeV}\cite{pdg}\\ \hline
			$\Gamma^5_{\pi\pi}$&$\Gamma[f_0(1710)\rightarrow\pi\pi]$& $(0.12 \pm 0.11)\times(220 \pm 40)$ {\rm MeV} \cite{oller}\\ \hline
			$\Gamma^5_{KK}$&$\Gamma[f_0(1710)\rightarrow K\bar{K}]$& $(0.36\pm 0.12)\times(220 \pm 40)$ {\rm MeV} \cite{oller}\\ \hline
			$\Gamma^5_{\eta\eta}$&$\Gamma[f_0(1710)\rightarrow \eta\eta]$& $(0.22 \pm 0.12)\times(220 \pm 40)$ {\rm MeV} \cite{oller} \\ \hline
		\end{tabular}
	\end{center}
	\caption{Target quantities used in global simulation III.
\label{quantities2}}
\end{table}
% % % % % %

Our objective is to explore the underlying mixings among various two-quark, four-quark and glue components in order to achieve a global understanding of all $I=0$ scalar states below 2 GeV.
This objective is sometimes at the expense of individual accuracies,  at least at the present approximation of the model.      Therefore,  we aim to determine the 14 Lagrangian parameters such that we get a reasonable overall agreement with all target inputs of Tables \ref{quantities1} and \ref{quantities2}.  In defining our numerical strategy we highlight three important points: (a) The experimental target quantities are of three different types (masses, decay widths and decay ratios); (b) since we are seeking a global understanding of isosinglet scalar mesons it is crucial for us to give the same importance to each targeted data regardless of whether they are of the same type or of different types; and (c) the fact that there are sometimes different reported experimental data for the same quantity (such as the cases discussed above), the role of a central value in a reported experimental data becomes less pronounced, and as a result, all points within a given experimental range become equally viable.     Either of points (a) or (b) rule out the use of conventional $\chi^2$ fits.   Instead,  we guide our numerical work by defining a function $\chi$
\begin{equation}
\chi\left( p_1\ldots p_{14} \right) = \sum_{i=1}^{N_q^\mathrm{exp}}
              \left|
              {
                { {\hat q}_i^\mathrm{exp} - q_i^\mathrm{theo}\left(p_1\ldots p_{14}\right)}
                               \over
                          {{\hat q}_i^\mathrm{exp}}
              }
               \right|,
\label{E_chi_def}
\end{equation}
where $q_i^\mathrm{exp} = {\hat q}_i^\mathrm{exp} \pm \Delta q_i^\mathrm{exp}$ are our target experimental quantities ($i=1\ldots N_q^\mathrm{exp}$),  which are also theoretically calculated by the model $q_i^\mathrm{theo}$ as a function of the 14 model parameters $p_1\ldots p_8$ = $c$, $d$, $c'$, $d'$, $m_G$, $\rho$, $e$, $f$ and $p_9\ldots p_{14}$ = $B$, $D$, $B'$, $D'$, $E$ and $F$ [for quantities that an experimental interval of the form $q_i^\mathrm{exp} = q_{i,\mathrm{min}}^\mathrm{exp} \rightarrow  q_{i,\mathrm{max}}^\mathrm{exp}$ is reported, we consider the target value to be the center  of the interval
${\hat q}_i^\mathrm{exp}=\left( q_{i,\mathrm{min}}^\mathrm{exp} + q_{i,\mathrm{max}}^\mathrm{exp} \right)/2$ and the uncertainty to be half the interval
$\Delta q_i^\mathrm{exp} =\left( q_{i,\mathrm{max}}^\mathrm{exp} - q_{i,\mathrm{min}}^\mathrm{exp} \right)/2$].   Clearly,  in the limit of $\chi\rightarrow 0$ the model predictions for target quantities approach the central values of their corresponding experimental data.   This limit, even if achievable,  is not of much physical significance due to the point (c) above.   Therefore, we do not misguide our computations by imposing this artificial limit.    To guide our computation we note that the overall experimental target quantities can be measured by
\begin{equation}
\chi^\mathrm{exp} =
\sum_{i=1}^{N_q^\mathrm{exp}}
              \left|
              {
                {\Delta q_i^\mathrm{exp}}
                               \over
                          {{\hat q}_i^\mathrm{exp}}
              }
               \right|.
\label{E_chiexp_def}
\end{equation}
Therefore,  to address point (c), we impose the condition
\begin{equation}
\chi \le \chi^\mathrm{exp},
\label{chi_best_generic}
\end{equation}
and filter out simulations that do not satisfy this condition.      Unlike the condition $\chi\rightarrow 0$ that results in finding a ``best point'' in a given simulation,  the final product of imposing condition (\ref{chi_best_generic}) is a set of acceptable points (each point is in the 14-dimensional parameter space of the model)\footnote{In other words,  although we aim at a set of ``bullseyes'' (central values of different experimental inputs), the criterion (\ref{chi_best_generic}) allows deviations from these ``bullseyes'' by measuring  the overall success (the global agreement between the model predictions and the central experimental values).}.    Once a parameter set is found, the physical quantities of interest (such as,  but not limited to,   all the input quantities, the quark and glue components of the isosinglet scalars, glueball mass, coupling constant, $\ldots$) are then determined at each point in the set and this process in turn results in a set of values for each quantity.

Specifically, our guiding function contains three parts
\begin{equation}
\chi \left(p_1\ldots p_{14}\right) = \chi_m \left(p_1\ldots p_8\right) + \chi_\Gamma \left(p_1\ldots p_{14}\right) +
\chi_{(\Gamma/\Gamma)} \left(p_1\ldots p_{14}\right),
\label{chi_generic}
\end{equation}
where the three terms on the right refer to mass, decay width and decay ratio, defined by
\begin{eqnarray}
\chi_m \left(p_1\ldots p_8\right) &=& \sum_{i=1}^5
              \left|
              {
                { {\hat m}_i^\mathrm{exp} - m_i^\mathrm{theo}\left(p_1\ldots p_8\right)}
                               \over
                          {{\hat m}_i^\mathrm{exp}}
              }
               \right|,
\nonumber \\
 \chi_\Gamma\left(p_1\ldots p_{14}\right) &=& \sum_{i=1}^5
              \sum_\alpha
              \left|
              {
                { \left({\hat \Gamma}_\alpha^i\right)^\mathrm{exp} -
                   \left(\Gamma_\alpha^i\right)^\mathrm{theo}\left(p_1\ldots p_{14}\right)}
                               \over
                          {\left({\hat \Gamma}_\alpha^i\right)^\mathrm{exp}}
              }
               \right|,
\nonumber \\
\chi_{(\Gamma/\Gamma)} \left(p_1\ldots p_{14}\right) &=& \sum_{i=1}^5
\sum_\alpha \sum_\beta
              \left|
              {
                { \left({\hat \Gamma^i_{\alpha/\beta}}\right)^\mathrm{exp} -
                \left(\Gamma^i_{\alpha/\beta}\right)^\mathrm{theo}\left(p_1\ldots p_{14}\right)}
                               \over
                          { {\left( {\hat \Gamma}^i_{\alpha/\beta}\right)}^\mathrm{exp}}
              }
               \right|,
\label{m_DW_DR_def}
\end{eqnarray}
with short notations
\begin{eqnarray}
\Gamma_\alpha^i &=& \Gamma \left[f_i \rightarrow \alpha\right], \nonumber \\
\Gamma^i_{\alpha/\beta} &=& { {\Gamma\left[f_i \rightarrow \alpha\right]}\over {\Gamma
\left[f_i \rightarrow \beta\right]}},
\end{eqnarray}
where $i=1\ldots 5$ correspond to the five isosinglet scalars in ascending order [Eq.~(\ref{K_def})],  $\alpha$ and $\beta$ are the two-body decay channels and in this work take values  $1\ldots 4$ which respectively correspond to the decay channels $\pi\pi$, $K\bar{K}$, $\eta \eta$ and $\eta \eta'$.

%%%%%%%%%%%%%%%%%%%%%%%%%%%%%%%%%%%%%%%%%%%%%%%%

\section{Results for simulation  I}

%%%%%%%%%%%%%%%%%%%%%%%%%%%%%%%%%%%%%%%%%%%%%%%%
In this section we focus on the target inputs given in Table \ref{quantities1} [with the exception of the two partial decay widths of $f_0(1370)$].   We first study the global simulation to determine  the first global set and then we impose additional constraints on the global set and study its two main subsets.     In next section, we give a comparison of the results obtained in this section with those obtained when all target inputs of Table \ref{quantities1} [including the partial decay widths of $f_0(1370)$] are taken into account; as well as when target inputs of Table \ref{quantities2} are used.

\subsection{Global picture: determining set $S_\mathrm{I}$}

In global fit I,  we exclude the partial decay widths of $f_0(1370)$ from our target experimental data of Table \ref{quantities1}.                                                                                                                    This means that the guiding function for this simulation ($\chi_\mathrm{I}$) is computed from (\ref{chi_generic})
\begin{equation}
\chi_\mathrm{I} \left(p_1\ldots p_{14}\right) = \chi_{\mathrm{I},m} \left(p_1\ldots p_8\right) + \chi_{\mathrm{I},\Gamma} \left(p_1\ldots p_{14}\right) +
\chi_{\mathrm{I},{(\Gamma/\Gamma)}} \left(p_1\ldots p_{14}\right),
\label{chi_I_def}
\end{equation}
 in which $\chi_{\mathrm{I},m}$, $\chi_{\mathrm{I},\Gamma}$ and $\chi_{\mathrm{I},(\Gamma/\Gamma)} $ are obtained from (\ref{m_DW_DR_def}),  with the condition that in $\chi_{\mathrm{I},\Gamma}$ the decay widths of $f_0(1370)$ have been excluded (i.e.\ $i\ne 3$).   The $\chi_{\mathrm{I},m}$ and $\chi_{\mathrm{I},(\Gamma/\Gamma)}$ are those given in (\ref{m_DW_DR_def}) and include all data for these two quantities given in Table \ref{quantities1}.  We use Monte Carlo simulation over the 14d parameter space and search for points $p=\left(p_1\ldots p_{14}\right)$ for which
\begin{equation}
\chi_\mathrm{I} (p) \le \chi_\mathrm{I}^\mathrm{exp},
\label{gfit_cond}
\end{equation}
subject to the constraint
\begin{equation}
\Gamma^3_{\pi\pi+KK+\eta\eta}=\Gamma[f_0(1370)\rightarrow \left(\pi\pi+ KK + \eta\eta\right)] < 500 \,\,\mathrm{MeV}.
\label{gfitI_const}
\end{equation}
In this case $\chi_\mathrm{I}^\mathrm{exp}=7.3$ (note that in computing $\chi_\mathrm{I}^\mathrm{exp}$  the central values of the experimental inputs $m_1$, $\Gamma^1_{\pi\pi}$ and $\Gamma^2_{\pi\pi}$ are considered, and that it receives no contribution from $m_3$, $m_4$ and $m_5$ that have fixed target values).  This leads to a set of points
\begin{equation}
S_\mathrm{I} = \left\{
 p \,| \, p\, \in \, \mathbb{R}^{14} \,\, : \,\, \chi_\mathrm{I} (p) \le \chi_\mathrm{I}^\mathrm{exp} \,\,\&\,\,  \Gamma^3_{\pi\pi+KK+\eta\eta} < 500 \,\,\mathrm{MeV} \right\}.
\label{S_I}
\end{equation}

Initially, the computation starts by generating random numbers for each of the 14 parameters over  relatively broad intervals.    It is then found that enforcing Eq.~(\ref{gfit_cond}) results in effectively limiting the range of variation of  each of these parameters into tighter regions.
This is shown in Fig.~\ref{F_raw_par} in which we see the finite range of variation of each of these parameters. Moreover, it is found that not all values of each parameter over its finite range of variation is acceptable (i.e.~the ranges of variation are not continuous).   This is due to the fact that the complicated guiding function $\chi_\mathrm{I}$ is a jagged function of the 14 parameters over their ranges of variation,  within which,  there are points that do not satisfy  condition (\ref{gfit_cond}).
%
% % % % % % % % % % % % % % % % % % % % figure 1
%
\begin{figure}
\begin{center}
\vskip .75cm
%-----------------------------------------
\epsfysize = 7cm
 \epsfbox{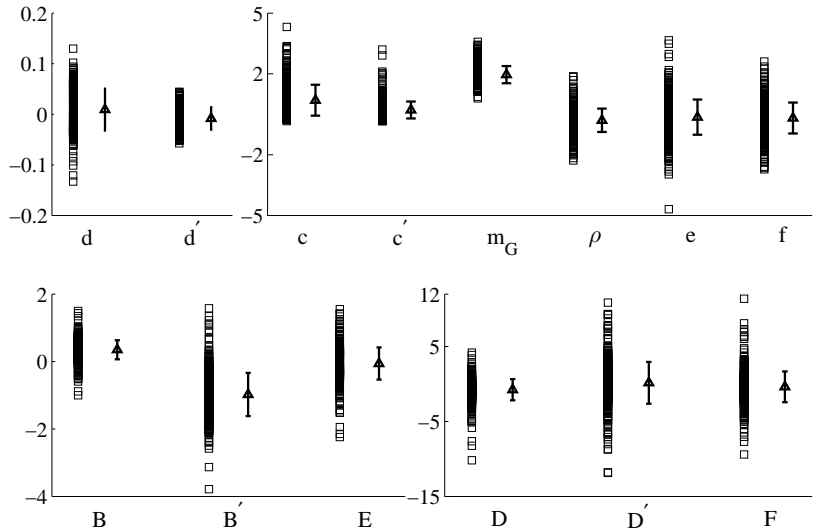}
%\hskip 1cm
\caption{The result of the Monte Carlo simulation (squares) over the fourteen Lagrangian parameters:  top eight ($c$, $d$, $\ldots$, $f$) define  the masses and mixing parts and bottom six ($B$, $D$, $\ldots$, $F$) define the  interaction part of the Lagrangian.    The points shown define set $S_\mathrm{I}$ [Eq.~(\ref{S_I})] at which the condition of Eq.~(\ref{gfit_cond}) is satisfied.   Also shown are the average values (triangles) and standard deviation around the averages (error bars).
\label{F_raw_par}}
\end{center}
\end{figure}

Figs.\ \ref{F_raw_masses}, \ref{F_raw_dws} and \ref{F_raw_drs} give the 21 input quantities  used in the global Monte Carlo simulation  I. After set $S_\mathrm{I}$ is determined,  these 21 quantities are then computed over $S_\mathrm{I}$ (i.e.~at every point in this set).   Also shown are the averages (triangles) and standard deviations around the averages (error bars).
The objective of the global fit I is to determine the set $S_\mathrm{I}$ such that the overall discrepancies between theory and experiment is no larger than the overall experimental uncertainties.    As a result,  some of the 21 quantities end up within their experimental ranges and some outside.    For example, in the case of masses shown in Fig.~\ref{F_raw_masses}, even though they span (or get very close to) the target values, but  we see that the sensitivity of simulations reduces in ascending order of mass.     This is roughly understandable since contribution of each of these masses to $\chi_{\mathrm{I},m}^\mathrm{exp}$ is inversely proportional to the mass and directly proportional to the experimental uncertainty [see Eq.~\ref{chi_generic}], hence a heavier state such as the $f_0(1710)$ with a larger mass and smaller experimental mass-uncertainty contributes much less to $\chi_{I,m}^\mathrm{exp}$ than the much lighter and less certain $f_0(500)$.    This leads to more smearing of simulations around $m[f_0(1710)]$ compared to that around $m[f_0(500)]$. The same property is observed for the  decay widths (Fig.~\ref{F_raw_dws}) and decay ratios (Fig.~\ref{F_raw_drs}).
%
% %  % % % % % % % % % % % % % % % % %% figure 2
%
\begin{figure}[h!]
\begin{center}
\vskip .75cm
%-----------------------------------------
\epsfysize = 7cm
 \epsfbox{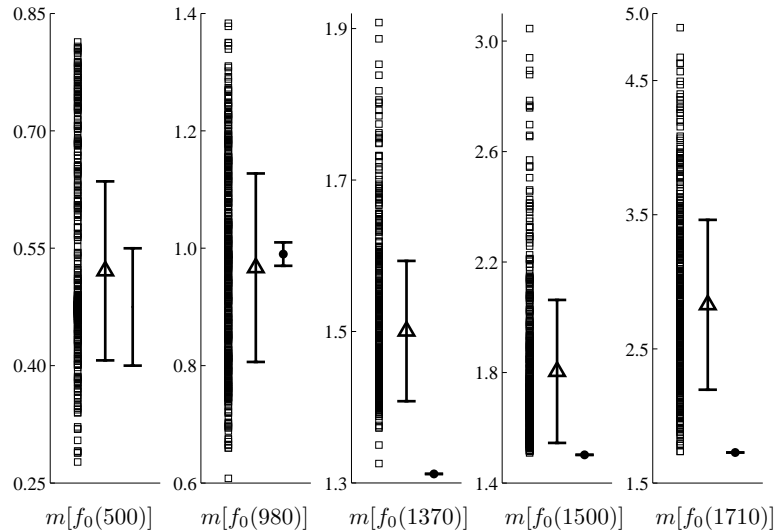}
%\hskip 1cm
\caption{Isosinglet scalar masses obtained in the global Monte Carlo simulation I (squares) over the 14 Lagrangian parameters  [set $S_\mathrm{I}$ defined in (\ref{S_I})] are compared with their experimental target ranges given in Table
\ref{quantities1} (solid circles). The results are obtained by solely requiring that the overall discrepancy between model predictions and experiment be no more than the overall experimental uncertainties. While simulations cover the acceptable values of all target masses,  their ranges of variation are considerably wider than the experimental uncertainties.   This suggests that additional filtering conditions (similar to those discussed in the subsequent subsections) are needed to further limit these ranges.  }
\label{F_raw_masses}
\end{center}
\end{figure}
%
% % % % % % % % % % % % % % % % % % % % figure 3
%
\begin{figure}[h]
\begin{center}
\vskip .75cm
%-----------------------------------------
\epsfysize = 7cm
 \epsfbox{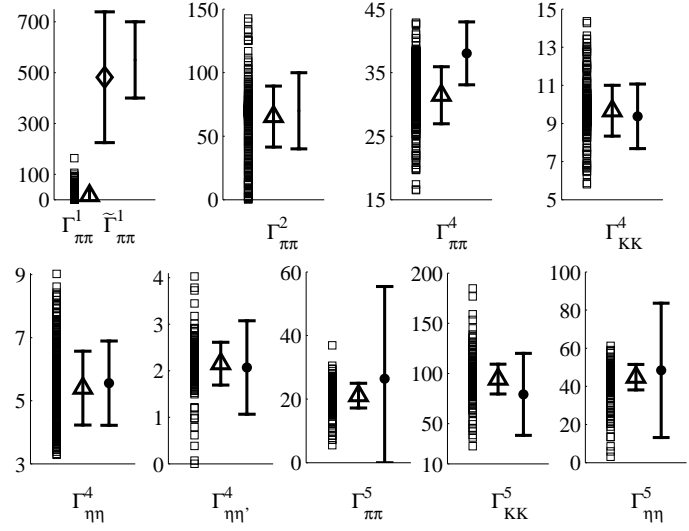}
%\hskip 1cm
\caption{Isosinglet scalar decay widths (squares) obtained in the Monte Carlo simulation I over the 14 Lagrangian parameters  are compared with their experimental target values (solid circles) of Table \ref{quantities1}.   For the case of decay width of $f_0(500)$ to two pions,  two computations are given:  The bare decay width $\left( \Gamma^1_{\pi\pi}\right)$ [squares, and their average and standard deviation (triangle and error bar)] as well as the  physical decay width $\left({\widetilde \Gamma}^1_{\pi\pi}\right)$ [average and standard deviation shown with diamond and error bar] in which the effect of final-state interactions of pions are taken into account.}
\label{F_raw_dws}
\end{center}
\end{figure}
%
% % %  % % % % % % % % % % % % % % % % % figure 4
%
\begin{figure}[h]
\begin{center}
\vskip .75cm
%-----------------------------------------
\epsfysize = 7cm
 \epsfbox{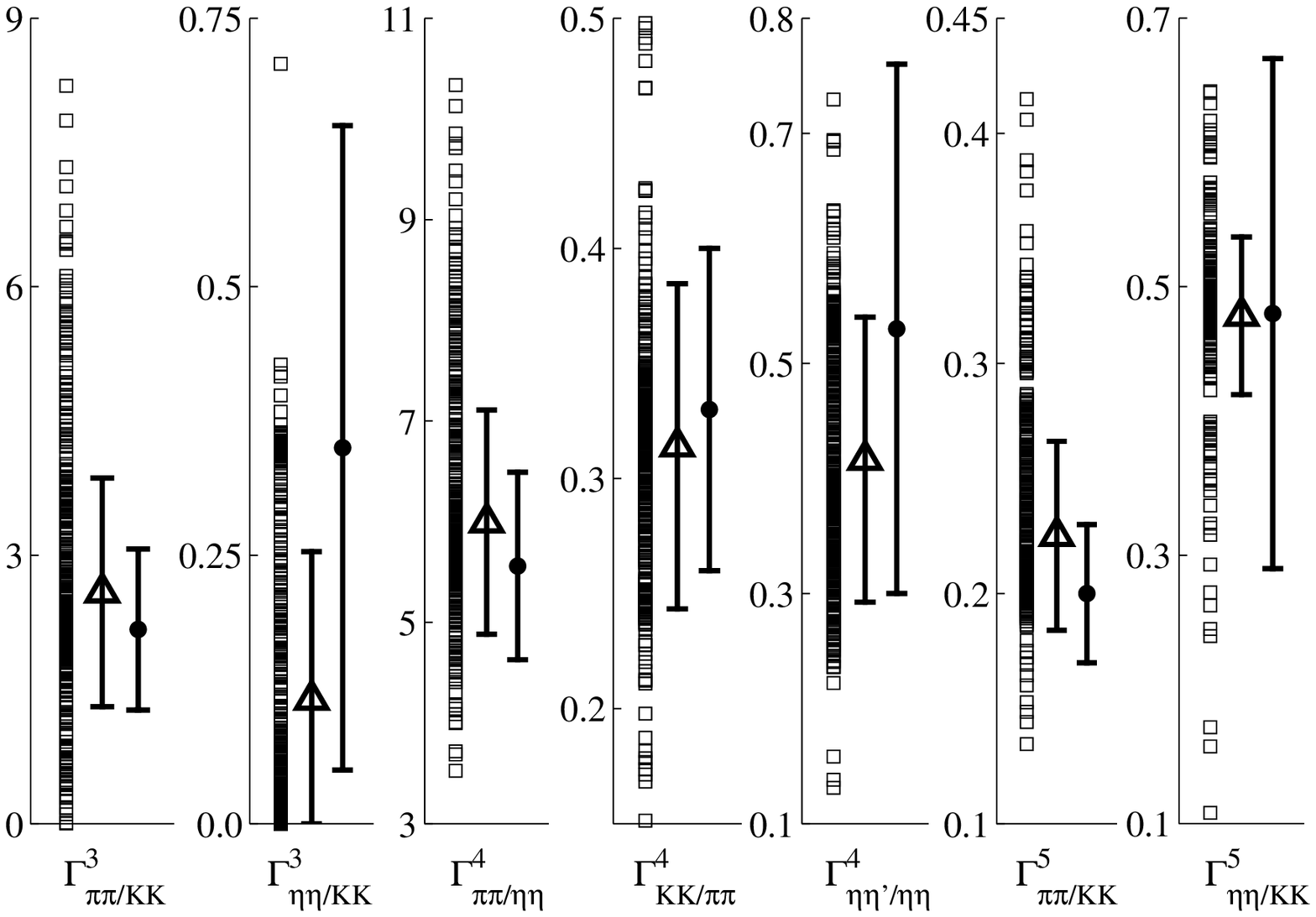}
%\hskip 1cm
\caption{Isosinglet scalar decay ratios (squares)  obtained in the Monte Carlo simulation over the 14 Lagrangian parameters  are compared with their experimental target values (solid circles) given in  Table \ref{quantities1}.
\label{F_raw_drs}}
\end{center}
\end{figure}

The smearing effects are not an artifact of the designed computational algorithm.  In fact they show that the algorithm is performing the computation correctly because of the following important point.     The $\chi$ method we use in aiming for the  central values of the experimental quantities,  assigns the same weight (i.e.\ gives the same importance) to all target quantities involved (i.e.\ the masses, the decay widths and the decay ratios displayed in Table \ref{quantities1}).  Compared to the conventional $\chi^2$ method,   the $\chi$ method (first introduced in \cite{06_F}) implements a democratic treatment of all target quantities regardless of their type or uncertainty.    However,
when it comes to implementing  the criterion (\ref{gfit_cond}) in order to allow deviations from central values,   the experimental quantities that are larger may naturally get overshadowed by smaller target quantities.    This highlights the tension between the global description and the local precision and that the inevitable price to pay for the former it to somewhat sacrifice the latter.   The value of the (less precise) global simulation is that it is based on fewer assumptions and that it explores family relations,  but obviously since there are no ``free lunches,''  it looses individual precisions. Nevertheless,  it forms a reliable first step in understanding a complicated system such as the isosinglet scalars that have tremendous underlying mixings.    Starting with a global simulation controls the overall ranges of model parameters and paves the way for zooming in on each individual state by imposing further refining conditions.

The results for the decay widths and decay ratios are shown in Figs.~\ref{F_raw_dws} and \ref{F_raw_drs} and we see that the simulation averages and standard deviations generally overlap with the experimental inputs. Note that for the decay width of $f_0(500)$ to two pions (the first subgraph) two computations are given.   The bare decay width of $f_0(500)$ to two pions [the simulation points (squares) and their average/standard deviation (triangle/error bar)] in which the effects of the final-state interactions of pions are not accounted for.   These effects are known to be important for a broad state such as $f_0(500)$ and have been calculated in this nonlinear chiral Lagrangian framework in description of $\pi\pi$ scattering in \cite{HSS1} in which it is shown that the physical decay width is considerably larger than the bare one.
Using our simulation data in set $S_\mathrm{I}$,  together with the methodology of \cite{HSS1},  we have added the effects of the final-state interactions of pions to this decay width  and plotted the average (diamond)  and the standard deviation (error bar) of the result in the same figure which then overlaps with the target range of 400--700 MeV \cite{pdg}.    (Further details on the numerical values of the physical  quantities of interest in simulation I, are given in Table \ref{T_App_B1} of Appendix B.)

The global simulation aims to reconcile the model parameters with the overall experimental inputs on the mass spectrum and decay properties,  and result in a large set of points in $S_\mathrm{I}$.  The quark and glue components of each of the five isosinglet scalars are computed over the global set $S_\mathrm{I}$ and presented in Fig.~\ref{F_raw_comps}.   While these points lead to predictions for the quark and glue components that spread over a wide range, some qualitative and average observations can be made.   With the exception of $f_0(1370)$ which is seen to exhibit significant  quark-antiquark components \cite{FAA_paper1},  for the other four states the global simulations clearly show that there is a significant underlying mixing among the two-quark, four-quark and glue components that form the scalar mesons.
Since our focus in this paper is on the substructure of $f_0(1500)$ and $f_0(1710)$ (particularly their glue content) we have organized the result of the global simulations for all isosinglet states in such a way that  correlations with the glue content of $f_0(1500)$  can be traced.    In Fig.~\ref{F_raw_comps}, for each of the isosinglet states,  the correlation between their five components and the glue component of $f_0(1500)$ [which is broken into four intervals 0--25\% ($\medstar$), 25\%--50\% ($\bigtimes$), 50\%--75\% ($\lozenge$) and 75\%--100\% ($\bigcircle$)] is shown.
For each component,  the raw data (dots) are also shown together with their averages and standard deviations; the four series of dots for each component from left to right correspond to simulations that have resulted the glue component of $f_0(1500)$  in the four ranges of 0--25\%, 25--50\%, etc.   The average and standard deviation for each series of dots is given to their right.    Although each four groups of dots for each component spans a wide range, the averages and standard deviations are stable for most of the components.   Exceptions are the four-quark components of $f_0(1500)$ and $f_0(1710)$ that are in a seesaw relationship with the glue components of these two states.   Also the glue components of $f_0(1500)$ and $f_0(1710)$ are in  a seesaw relationship, which is the most clear signal for the large glue component of these two scalar mesons in this global simulation.  Note that this observation is merely the direct result of confronting the model with the experimental inputs and  without any additional conditions imposed.         Beyond the fact that the raw simulation data shows that the glue is almost exclusively shared between $f_0(1500)$ and $f_0(1710)$, it is nontrivial to gain  further detailed insight into the glue percentages  of each of these two states at this level of global computation.      For convenience the sum of the two two-quark and the two four-quark components are also displayed on the right in Fig.~\ref{F_raw_comps}.

Several comments are in order:
Both the $f_0(500)$ and the $f_0(980)$ have a low glue component, but a significant mixture of two- and four-quark components.  The four-quark component of $f_0(500)$ is dominant consistent with the expected four-quark nature of this scalar meson suggested in MIT bag model \cite{Jaf} and elsewhere.     The content of $f_0(980)$ is rather reversed.  However,  since the range of variation of the two and four-quark components of the $f_0(500)$ and the $f_0(980)$  are wide,  further evaluation of this raw simulation is needed and will be considered next.
The $f_0(1370)$ has the most stable content with dominant  two-quark components; in this case the ranges of simulations are much narrower than the other four states (for a detail study of $f_0(1370)$ within this framework see \cite{FAA_paper1}).
Lack of a significant glue component for $f_0(500)$, $f_0(980)$  and $f_0(1370)$ leaves the main competition for this component between $f_0(1500)$ and $f_0(1710)$.    In order to identify which of these two are the main glue holders, we need to further analyze the raw data and impose several additional conditions on the data and try to see if we can  filter out some of the raw data.   These will be done in the following subsections.
The overall averages are given in Table \ref{T_GFI_comps} and
show that the $f_0(1500)$ has the dominant glue average followed by that of $f_0(1710)$.  However,  due to the overlapping distribution of simulations  around these two glue averages it becomes a challenging problem to pin point exactly which of these two states has the dominant glue.
\begin{table}[h]
\begin{center}
	\begin{tabular}{|c||c|c|c|c|c||c|c|}
		\hline
		&	${\bar u}{\bar d} u d$ & $\frac{{\bar d}{\bar s} d s + {\bar s}{\bar u} s u}{\sqrt{2}}$ & $s {\bar s}$ & 	 $\frac{u {\bar u}  + d {\bar d}}{\sqrt{2}}$ &  $G$ & Total four-quark & Total two-quark  \\ \hline
			
		$ f_0(500)  $  & 	$ 42  $  &  $ 22   $  &  $ 15   $  &  $ 18   $  &  $  3   $ & $ 64   $  &  $ 33   $ \\ \hline
		$ f_0(980)  $  & 	$ 22  $  &  $ 13   $  &  $ 27   $  &  $ 30   $  &  $  8   $ & $ 35   $  &  $ 57   $ \\ \hline
		$ f_0(1370) $  & 	$  8  $  &  $  6   $  &  $ 49   $  &  $ 33   $  &  $  4   $ & $ 14   $  &  $ 82   $ \\ \hline
		$ f_0(1500) $  & 	$ 10  $  &  $ 23   $  &  $  3   $  &  $  4   $  &  $ 60   $ & $ 33   $  &  $  7   $ \\ \hline
		$ f_0(1710) $  & 	$ 18  $  &  $ 36   $  &  $  6   $  &  $ 15   $  &  $ 25   $ & $ 54   $  &  $ 21   $ \\ \hline
\end{tabular}
\end{center}
\caption{The percentages of components averaged over set $S_\mathrm{I}$ [definition (\ref{S_I})] obtained in global Monte Carlo simulation I.}
\label{T_GFI_comps}
\end{table}

Before we end this subsection,  for convenience,  we further dissect the results of this global simulation by presenting the components in histograms.     This is shown in Fig.~\ref{F_raw_histogram} where the distribution of components of each isosinglet state at points in $S_\mathrm{I}$ is given over a five-interval division. Consistent with the above observations,   the $f_0(1500)$ shows the maximum distribution over the large glue intervals (80--100\%) followed by the glue of $f_0(1710)$.   It is also seen that the four-quark components of $f_0(500)$ and $f_0(980)$ are large with a considerable mixing with their two-quark components.    Again the $f_0(1370)$ appears to be the most clear cut case with dominant quark-antiquark components.

% % % % % % % % % % % % % % % % % % % % figure 5

\begin{figure}[h]
\begin{center}
%-----------------------------------------
\epsfxsize = 17cm
 \epsfbox{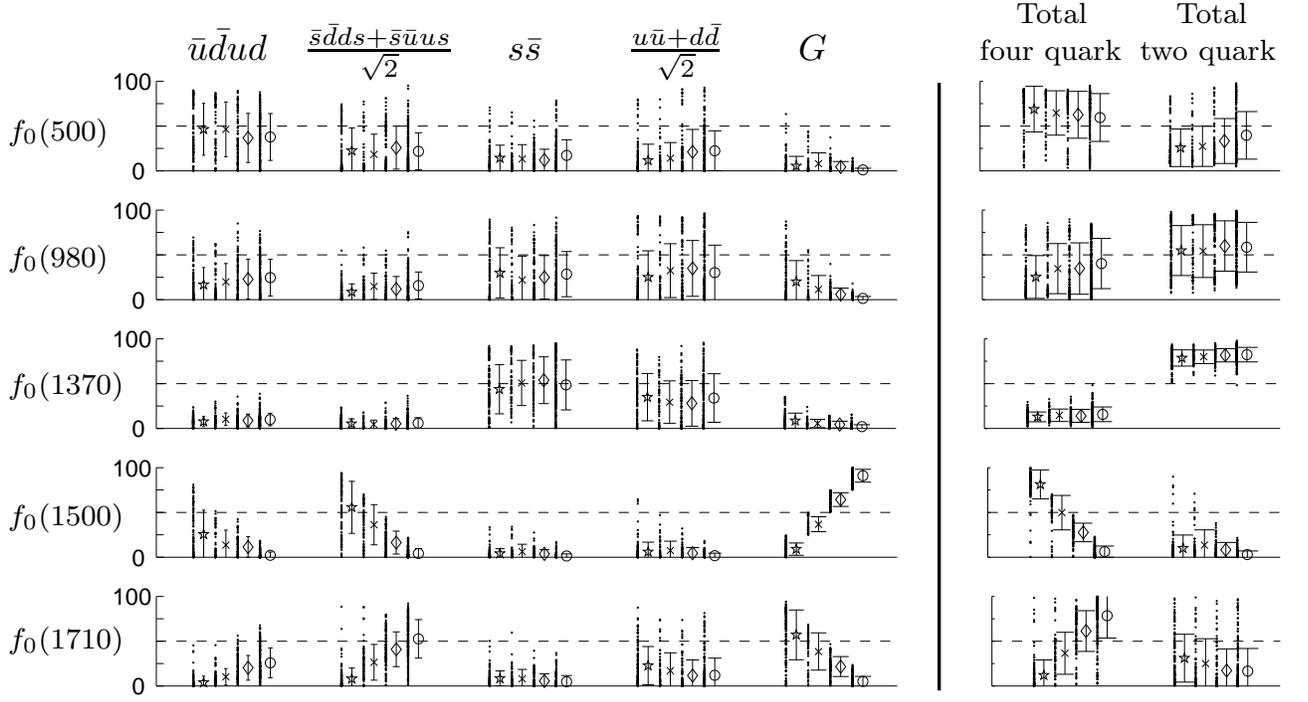}
\caption{Correlation diagrams for the quark and glue components of the isosinglet scalars obtained in the global Monte Carlo simulation I over the 14 Lagrangian parameters.   In the left figure, from left to right the five components are respectively non-strange four-quark, strange four-quark, strange quark-antiquark, non-strange quark-antiquark and glue.    In the right figure, for convenience of comparison, the sum of the two four-quark components and the sum of the two quark-antiquark components are shown.  The correlation between each component and the glue content of $f_0(1500)$ are shown in the following manner:  The overall simulations for each component is divided into  four groups [four vertical lines of dots, each dot is a computed component at a point in set $S_\mathrm{I}$ of definition (\ref{S_I})]; the vertical lines of dots from left to right correspond to simulations for which the glue content of $f_0(1500)$ is in the ranges  0--25\%, 25--50\%, 50--75\% and 75--100\%, respectively (see the glue content of $f_0(1500)$, for a definition of these four groups).  Next to each vertical line of dots the averages (symbols) and one standard deviation around the averages (error bars) are also shown.    In both figures, for convenience,  the 50\% lines are plotted as well (dashed lines). The ranges of variation reflect  $\chi_\mathrm{I} <\chi_\mathrm{I}^\mathrm{exp}$ condition defined in (\ref{gfit_cond}).}
\label{F_raw_comps}
\end{center}
\end{figure}

% % % % % % % % % % % % % % % % % % % % figure 6

\begin{figure}[h]
\begin{center}
\vskip .75cm
%-----------------------------------------
\epsfxsize = 17cm
 \epsfbox{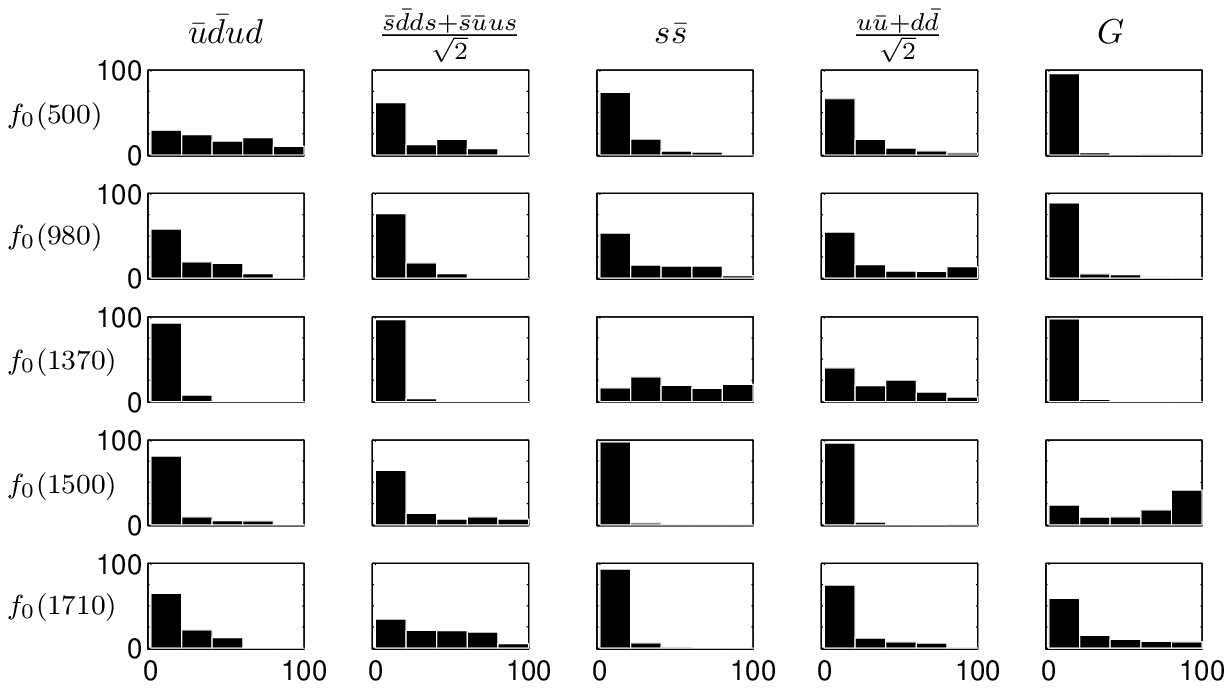}
%\hskip 1cm
\caption{Histograms for quark and glue components of the isosinglet scalars obtained in the global Monte Carlo simulation I over the 14 Lagrangian parameters.   Each histogram gives the percentage of the global simulation I (i.e.\ percentage of points in $S_\mathrm{I}$) over the five-interval breakdown of each component (0--20\%, 20\%--40\%, $\ldots$).    For example,  we see that respectively 29\%, 24\%, 16\%, 21\% and 10\% of simulations have resulted the ${\bar u}{\bar d}ud$ component of $f_0(500)$ in ranges 0--20\%, 20\%--40\%, etc.
Clearly,  $f_0(500)$ and $f_0(980)$ have considerable mixing of two- and four-quark components  with negligible glue; the $f_0(1370)$ is dominantly a two-quark state; $f_0(1500)$ contains the highest glue component followed by the glue component of $f_0(1710)$.
\label{F_raw_histogram} }
\end{center}
\end{figure}

%***********************************

\subsection{Imposing constrains on the global set $S_\mathrm{I}$}

%***********************************

In previous subsection, we determined the global set $S_\mathrm{I}$ that contains a large set of points in the 14d parameter space of the model at which the model generally overlaps with the target inputs of Table \ref{quantities1} [excluding the two decay widths of $f_0(1370)$].     This global set also allowed making average predictions for the substructure of isosinglet states.   In this subsection we further explore these substructures by imposing several additional constraints on set $S_\mathrm{I}$ that results in filtering out some of the points in that set.

We noted in Fig.~\ref{F_raw_masses} that the masses of $f_0(1500)$ and $f_0(1710)$ have a wide range in global set $S_\mathrm{I}$.  Even though the physical target masses are nearly included in this set but a large part of the set is far from the physical masses and we filter them out here.  As stated before, the WA102 \cite{WA102} has set the masses of $f_0(1370)$, $f_0(1500)$ and $f_0(1710)$ to fixed values which are not exactly recovered in the global set $S_\mathrm{I}$, but we are able to get close to them within about 5\% error (the mass ranges reported in other sources \cite{pdg,oller} have uncertainties, particularly $f_0(1370)$ is reported in PDG \cite{pdg} with a large mass uncertainty in the range of 1.2--1.5 GeV). We define subset $S_{\mathrm{I}1}$ such that the mass of $f_0(1710)$ is within 5\% of its  value fixed in WA102 collaboration report (since the predicted masses are sorted in ascending order, imposing a limit on $f_0(1710)$ mass naturally brings the mass of $f_0(1500)$ near its target value):
\begin{equation}
S_{\mathrm{I}1} = \left\{
 p \,| \, p\, \in \, S_\mathrm{I} \,\, : \,\, m[f_0(1710)]=1727\pm 86 \,{\rm MeV}\right\}.
\label{S_I1}
\end{equation}
Selected sample points in subset (\ref{S_I1}) are given in
Table \ref{T_App_A_SI1} in Appendix \ref{app1}.
The resulting averages (and standard deviations) for the masses,  decay widths and decay ratios are computed over subset $S_{\mathrm{I}1}$ and plotted in Fig.~\ref{F_mfilter_results}.    The close overlaps with the target input values of Table \ref{quantities1} is evident.   We then compute the quark and glue components of isosinglet states over subset $S_{\mathrm{I}1}$ and the results are presented in Fig.~\ref{F_mfilter_comps}.   We see that imposing the mass condition does not significantly change the qualitative description of the isosinglet states.   The $f_0(500)$ and $f_0(980)$ have a substantial two- and four-quark mixing;  the $f_0(1370)$ remains a predominantly quark-antiquark state; and the $f_0(1500)$ and $f_0(1710)$ remain to be the main glue holders and in this case the glue is almost equally shared among these two states.
%
% %  % % % % % % % % % % % % % % % % %% figure 2
%
\begin{figure}[h!]
\begin{center}
%\vskip .75cm
%-----------------------------------------
\epsfysize = 4.2cm
 \epsfbox{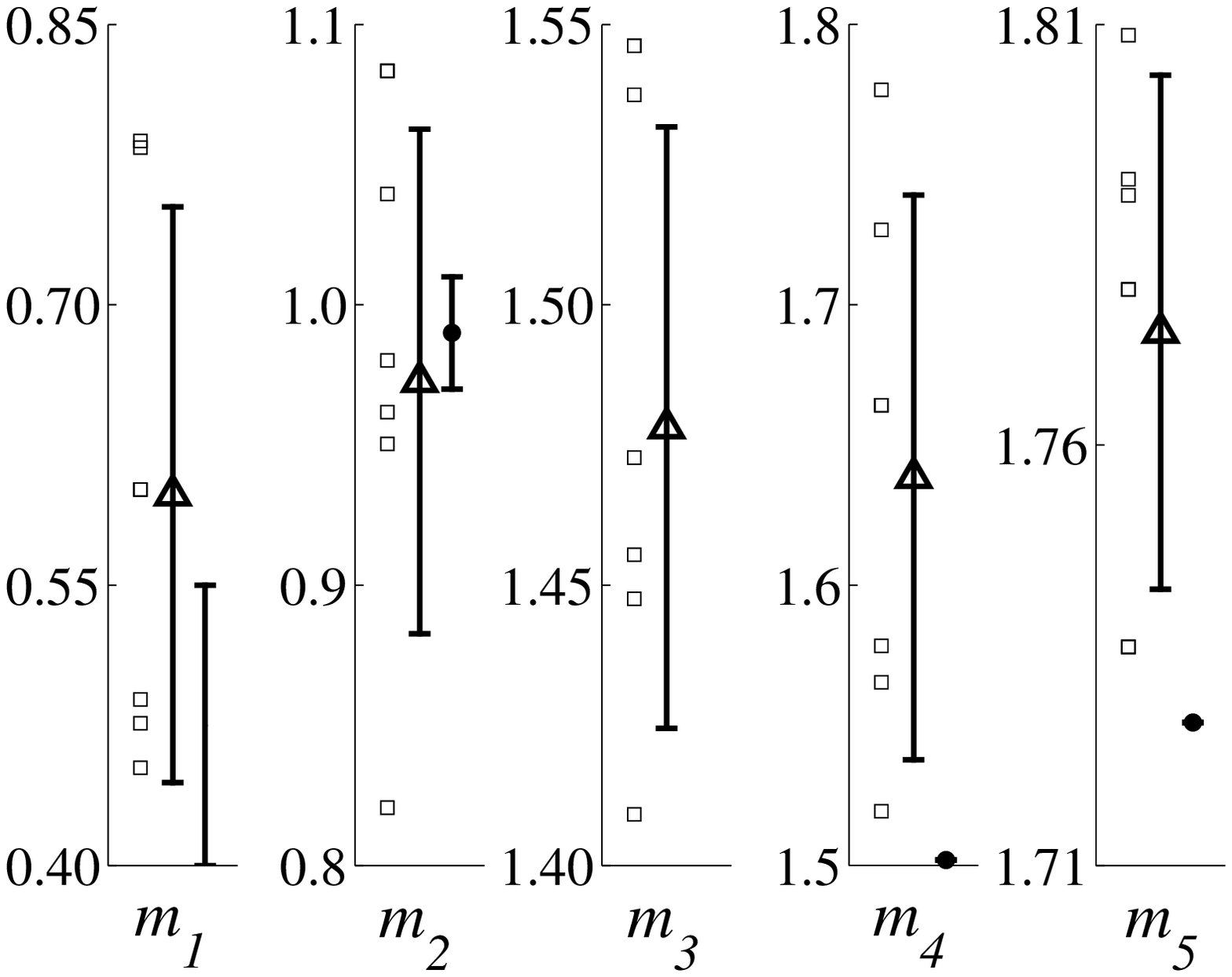}
 \hskip .6cm
\epsfysize = 4.2cm
 \epsfbox{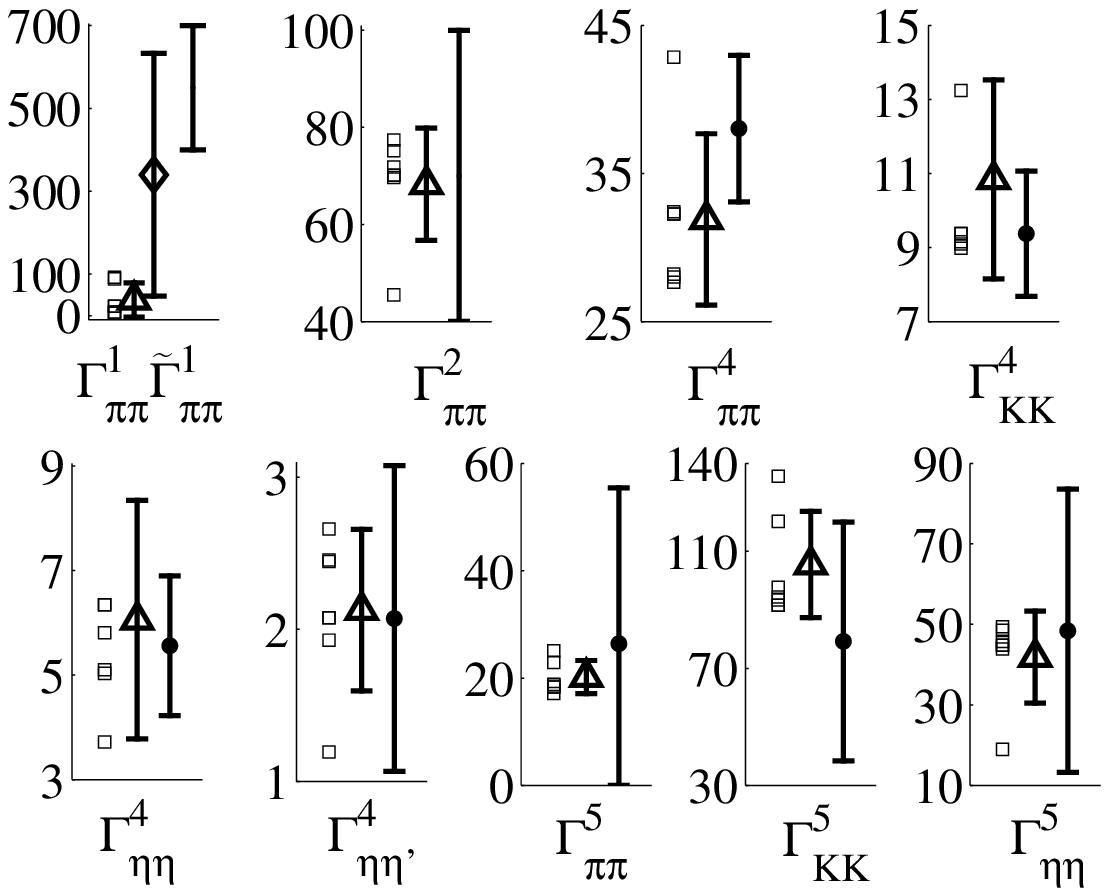}
\hskip .7cm
\epsfysize = 4.2cm
 \epsfbox{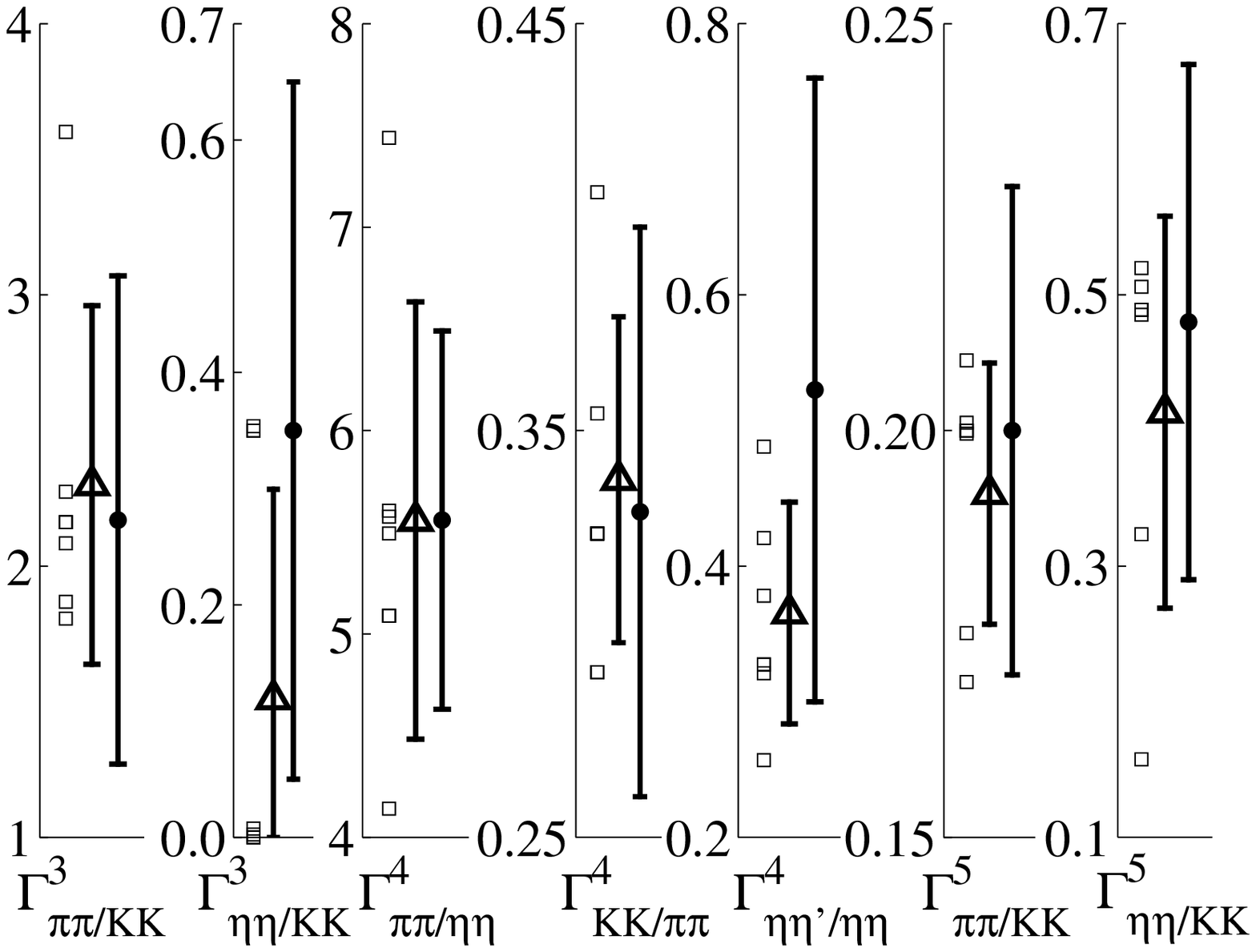}
%\hskip 1cm
\caption{Model predictions [averages(triangles) and standard deviations (error bars)]  for masses (left) decay widths (middle) and decay ratios (right) are compared with corresponding  inputs of Table \ref{quantities1} (solid circles and error bars).    The predictions are made over the subset $S_{\mathrm{I}1}$ defined in  (\ref{S_I1}).
 }
\label{F_mfilter_results}
\end{center}
\end{figure}
\begin{figure}[h]
\begin{center}
%\vskip .75cm
%-----------------------------------------
\epsfxsize = 17cm
 \epsfbox{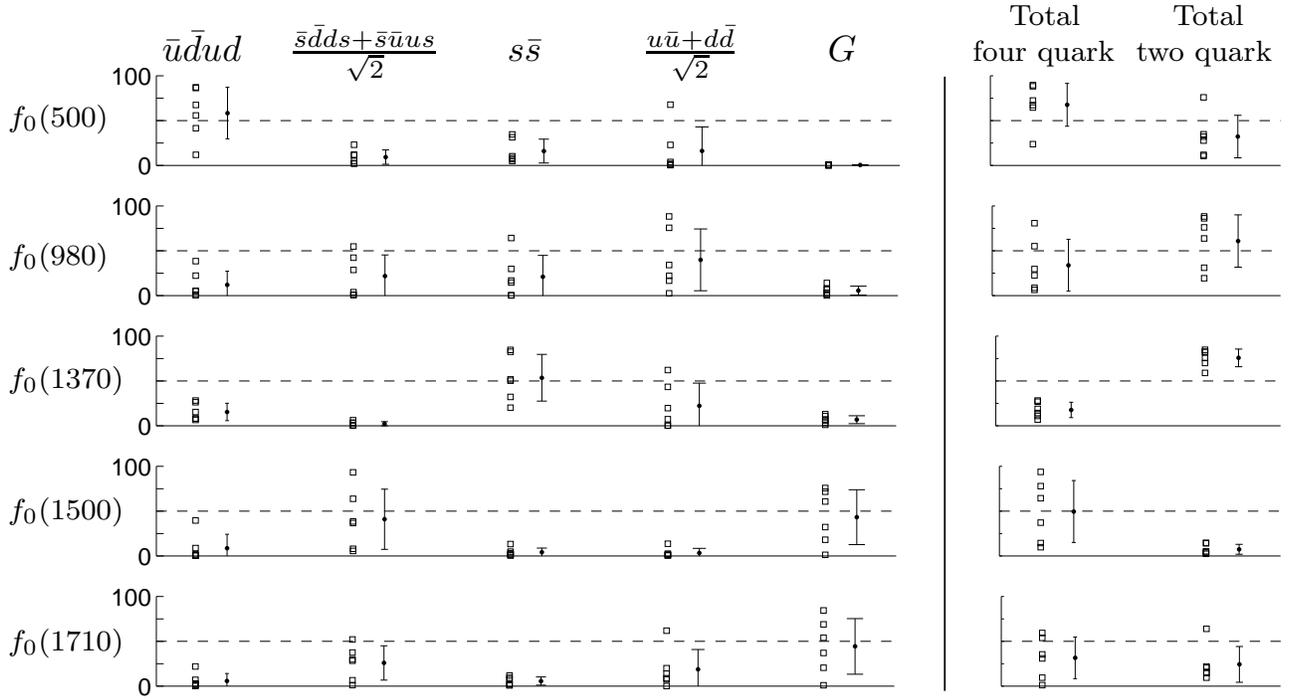}
\caption{Quark and glue components of the isosinglet scalars computed
over the subset $S_{\mathrm{I}1}$ [defined in (\ref{S_I1})].
In the left figure, from left to right the five components are respectively non-strange four-quark, strange four-quark, strange quark-antiquark, non-strange quark-antiquark and glue.    In the right figure, the sum of the two four-quark components and the sum of the two quark-antiquark components are shown.  For convenience,  the 50\% lines are also shown (dashed lines).
}
\label{F_mfilter_comps}
\end{center}
\end{figure}

\begin{table}[h]
	\begin{center}
		\begin{tabular}{|c||c|c|c|c|c||c|c|}
			\hline
		&	${\bar u}{\bar d} u d$ & $\frac{{\bar d}{\bar s} d s + {\bar s}{\bar u} s u}{\sqrt{2}}$ & $s {\bar s}$ & 	 $\frac{u {\bar u}  + d {\bar d}}{\sqrt{2}}$ &  $G$ & Total four-quark & Total two-quark  \\ \hline
			
		$ f_0(500)  $  & 	$ 58  $  &  $  9   $  &  $ 17   $  &  $ 16   $  &  $  0  $ & $ 67   $  &  $ 33   $ \\ \hline
		$ f_0(980)  $  &	$ 12  $  &  $  22  $  &  $ 20   $  &  $ 40   $  &  $  6  $ & $ 34   $  &  $ 60   $ \\ \hline
		$ f_0(1370) $  &	$ 15  $  &  $  2   $  &  $ 54   $  &  $ 22   $  &  $  7  $ & $ 17   $  &  $ 76   $ \\ \hline
		$ f_0(1500) $  &	$  9  $  &  $ 41   $  &  $  4   $  &  $  3   $  &  $ 43  $ & $ 50   $  &  $  7   $ \\ \hline
		$ f_0(1710) $  &	$  6  $  &  $ 26   $  &  $  5   $  &  $ 19   $  &  $ 44  $ & $ 32   $  &  $ 24   $ \\ \hline
		\end{tabular}
	\end{center}
\caption{The percentages of components averaged over subset $S_{\mathrm{I}1}$ defined in (\ref{S_I1}).}
\label{T_mfilter_comps}	
\end{table}

The global picture considered in this work has its  downside and upside.   The downside, first and foremost,  is the fact that dealing with all scalar states below 2 GeV at the same time, expectedly,  results in exceeding complications.   Second,  we loose precision on individual states in order to determine a set of parameters that give a  collective description of all states in an overall agreement with experiment.   The upside, however,  is that the global picture allows establishing underlying correlations (or family relations) among different states which can then be used to probe fuzzy situations such as exploring the glue content of $f_0(1500)$ and $f_0(1710)$.   For example,  the expectations that  $f_0(500)$ and $f_0(980)$ do not have significant glue contents (which seems to be supported by most, if not all, independent investigations), can be used as a filter to be imposed on set $S_\mathrm{I}$.
Fig.~\ref{F_raw_comps} shows the correlation between glue of $f_0(980)$ and those of $f_0(1500)$ and $f_0(1710)$.   If we turn off the glue of $f_0(500)$ and $f_0(980)$ (i.e.\ symbols ``$\medstar$'' and ``$\bigtimes$''), then the $f_0(1500)$ becomes the state with main glue component followed by $f_0(1710)$.
We further tap into  correlations between the substructure of the $f_0(500)$ and $f_0(980)$ on the one hand, and the $f_0(1500)$ and $f_0(1710)$, on the other. Specifically,    we filer the global set $S_\mathrm{I}$ so that  the $f_0(500)$ and $f_0(980)$ approach the picture given by the MIT bag model  \cite{Jaf} in which the lowest-lying scalar meson nonet is an ideally mixed two-quark two-antiquark nonet, in which
\begin{eqnarray}
f_0(500) &\propto& {\bar u}{\bar d} u d,  \nonumber\\
f_0(980) &\propto& { { {\bar d}{\bar s} d s + {\bar s}{\bar u} s u}\over {\sqrt{2}}}.
\end{eqnarray}
Since our model includes various mixings among isosinglets, we cannot exactly get to this limit, however,  we can approach it by imposing the condition  that the first component of $f_0(500)$ and the second component of $f_0(980)$, defined in (\ref{SNS_def}),  are each greater than 0.50, i.e.\ we filter the global set $S_\mathrm{I}$ with the condition
\begin{eqnarray}
\left(K_{11}^{-1}\right)^2 & > & 0.50, \nonumber\\
\left(K_{22}^{-1}\right)^2 & > & 0.50.
\label{ideal_cond}
\end{eqnarray}
This condition defines the second subset:
\begin{equation}
S_{\mathrm{I}2} = \left\{
 p \,| \, p\, \in \, S_\mathrm{I} \,\,: \,\,
 \left(K_{11}^{-1}\right)^2  >  0.50 \,\, \&\,\,
 \left(K_{22}^{-1}\right)^2  >  0.50 \right\}.
\label{S_I2}
\end{equation}

We find that this condition retains only a limited number of points in set $S_\mathrm{I}$.   The quark and glue components are then computed at these points and the result is given in Fig.~\ref{F_4q_comps}.   It is evident from both figures that approaching an ideally mixed four-quark limit for lowest-lying scalars is consistent with a dominant glue component in $f_0(1500)$.
%
% % % % % % % % % % % % % % % % % % % % figure 8
%
\begin{figure}[h]
\begin{center}
%\vskip .75cm
%-----------------------------------------
\epsfxsize = 17cm
 \epsfbox{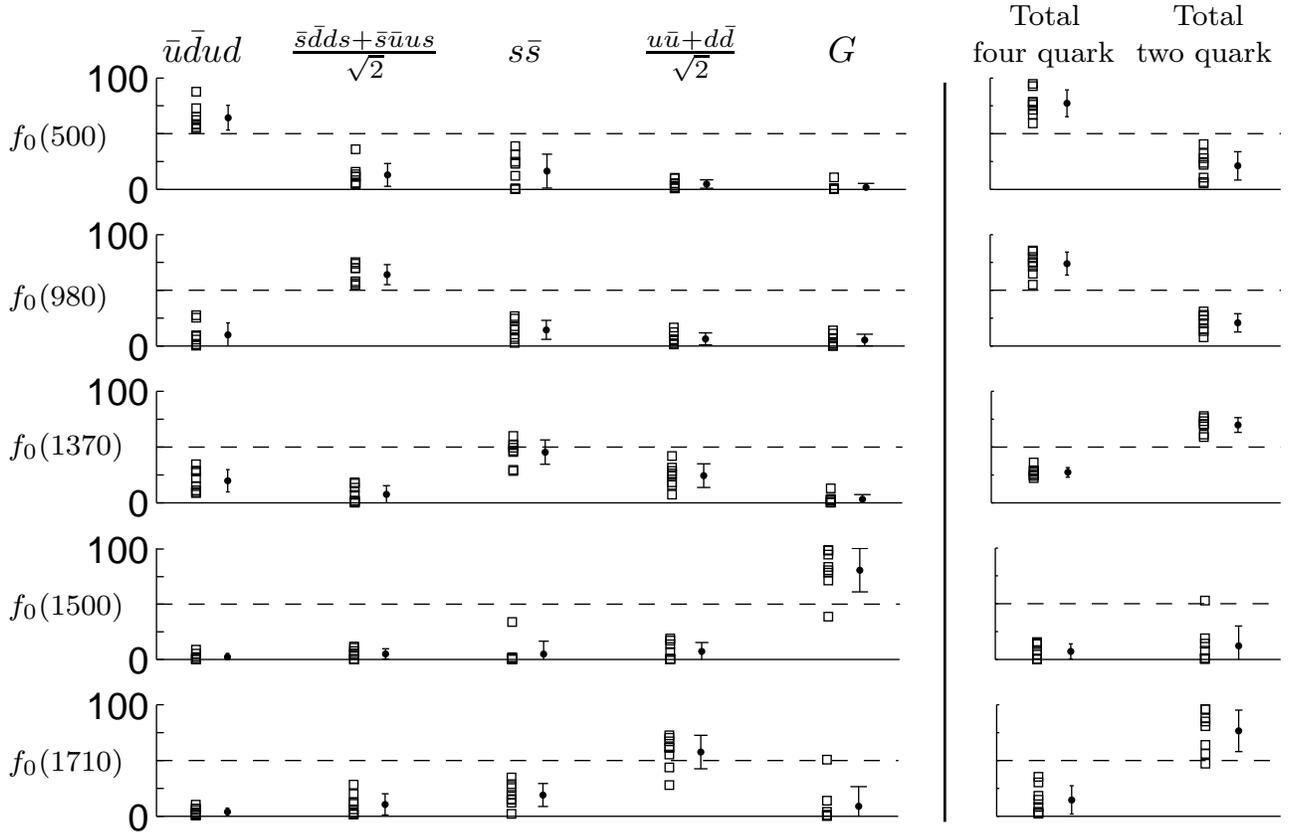}
%\hskip 1cm
\caption{Quark and glue components of the isosinglet scalars computed over  the subset $S_{\mathrm{I}2}$ [defined in (\ref{S_I2})] are shown on the left.     The Monte Carlo simulation is directed to approach a limit where $f_0(500)$ and $f_0(980)$ get close to their ideal mixing in a  four-quark scalar meson nonet [see (\ref{ideal_cond})]. For comparison, on the right,  the sum of the two four-quark components and the sum of the two quark-antiquark components are given.  For convenience,  the 50\% lines are also shown (dashed lines).}
\label{F_4q_comps}
\end{center}
\end{figure}

\begin{table}[h]
	\begin{center}
		\begin{tabular}{|c||c|c|c|c|c||c|c|}
			\hline
		&	${\bar u}{\bar d} u d$ & $\frac{{\bar d}{\bar s} d s + {\bar s}{\bar u} s u}{\sqrt{2}}$ & $s {\bar s}$ & 	 $\frac{u {\bar u}  + d {\bar d}}{\sqrt{2}}$ &  $G$ & Total four-quark & Total two-quark  \\ \hline
			
		$ f_0(500)  $  &	$ 64  $  &  $ 13   $  &  $ 16   $  &  $  5   $  &  $  2   $ & $ 77   $  &  $ 21   $ \\ \hline
		$ f_0(980)  $  &	$ 10  $  &  $ 64   $  &  $ 15   $  &  $  6   $  &  $  5   $ & $ 74   $  &  $ 21   $ \\ \hline
		$ f_0(1370) $  &	$ 20  $  &  $  8   $  &  $ 45   $  &  $ 24   $  &  $  3   $ & $ 28   $  &  $ 69   $ \\ \hline
		$ f_0(1500) $  &	$  2  $  &  $  5   $  &  $  5   $  &  $  7   $  &  $ 81   $ & $  7   $  &  $ 12   $ \\ \hline
		$ f_0(1710) $  &	$  4  $  &  $ 10   $  &  $ 19   $  &  $ 58   $  &  $  9   $ & $ 14   $  &  $ 77   $ \\ \hline
		\end{tabular}
	\end{center}
\caption{The percentages of components averaged over subset $S_{\mathrm{I}2}$ defined in (\ref{S_I2}).}
\label{T_ifilter_comps}	
\end{table}

%%%%%%%%%%%%%%%%%%%%%%%%%%%%%%%%%%%%%%%%%%%%%%%%

\section{Sensitivity to the experimental inputs:  Fits II and III}

%%%%%%%%%%%%%%%%%%%%%%%%%%%%%%%%%%%%%%%%%%%%%%%%

The results of the global simulation I presented in previous section were based on targeting the experimental inputs given in Table \ref{quantities1} [with the exception of the two decay widths of $f_0(1370)$].   Since the experiment is not firmly established on some of these inputs, as we discussed in section III, we have not based our entire analysis  on only one set of target experimental inputs.    In this section we present the main results we obtain in global simulation II [in which all target experimental inputs of Table \ref{quantities1}, including the two decay widths of $f_0(1370)$ are taken into account] as well as global fit III (in which the target experimental inputs of Table \ref{quantities2} are used).

In global fit II,  we target all 23 experimental data of Table \ref{quantities1}.                                                                                                                    This means that the guiding function for this strategy ($\chi_\mathrm{II}$) is computed from (\ref{chi_generic})
\begin{equation}
\chi_\mathrm{II} \left(p_1\ldots p_{14}\right) = \chi_{\mathrm{II},m} \left(p_1\ldots p_8\right) + \chi_{\mathrm{II},\Gamma} \left(p_1\ldots p_{14}\right) +
\chi_{\mathrm{II},{(\Gamma/\Gamma)}} \left(p_1\ldots p_{14}\right),
\label{chi_II_def}
\end{equation}
 in which $\chi_{\mathrm{II},m}$, $\chi_{\mathrm{II},\Gamma}$ and $\chi_{\mathrm{II},(\Gamma/\Gamma)} $ are obtained from (\ref{m_DW_DR_def}),  with all data of Table \ref{quantities1}.
We perform  Monte Carlo simulation over the 14d parameter space and search for points $p=\left(p_1\ldots p_{14}\right)$ for which
\begin{equation}
\chi_\mathrm{II} (p) \le \chi_\mathrm{II}^\mathrm{exp},
\label{SII_cond}
\end{equation}
In this case $\chi_\mathrm{II}^\mathrm{exp}=8.2$. This leads to  set II
\begin{equation}
S_\mathrm{II} = \left\{
 p \,| \, p\, \in \, \mathbb{R}^{14} \,\, : \,\,  \chi_\mathrm{II} (p) \le \chi_\mathrm{II}^\mathrm{exp} \right\}.
\label{S_II}
\end{equation}
We also explore two subsets of $S_\mathrm{II}$.   Imposing constrains on the mass of $f_0(1710)$ results in subset 1:
\begin{equation}
S_{\mathrm{II}1} = \left\{
 p \,| \, p\, \in \, S_\mathrm{II} \,\, : \,\, m[f_0(1710)]=1727\pm 86 \,{\rm MeV}\right\}.
\label{S_II1}
\end{equation}
Selected sample points in subset (\ref{S_II1}) are given in
Table \ref{T_App_A_SII1} in Appendix \ref{app1}.
Approaching the ideal mixing limit (\ref{ideal_cond}) leads to subset 2:
\begin{equation}
S_{\mathrm{II}2} = \left\{
 p \,| \, p\, \in \, S_\mathrm{II} \,\,: \,\,
 \left(K_{11}^{-1}\right)^2  >  0.50 \,\, \&\,\,
 \left(K_{22}^{-1}\right)^2  >  0.50 \right\}.
\label{S_II2}
\end{equation}

In global fit III,  we target the experimental data of Table \ref{quantities2}. The guiding function for this case ($\chi_\mathrm{III}$) is computed from (\ref{chi_generic})
\begin{equation}
\chi_\mathrm{III} \left(p_1\ldots p_{14}\right) = \chi_{\mathrm{III},m} \left(p_1\ldots p_8\right) + \chi_{\mathrm{III},\Gamma} \left(p_1\ldots p_{14}\right),
\label{chi_III_def}
\end{equation}
 in which $\chi_{\mathrm{III},m}$ and $\chi_{\mathrm{III},\Gamma}$  are obtained from (\ref{m_DW_DR_def}),  with all data of Table \ref{quantities2}.
With  Monte Carlo simulation over the 14d parameter space and searches for points $p=\left(p_1\ldots p_{14}\right)$ for which
\begin{equation}
\chi_\mathrm{III} (p) \le \chi_\mathrm{III}^\mathrm{exp},
\label{S_III_cond}
\end{equation}
with $\chi_\mathrm{III}^\mathrm{exp}=5.6$, results in set III
\begin{equation}
S_\mathrm{III} = \left\{
 p \,| \, p\, \in \, \mathbb{R}^{14} \,\, : \,\,   \chi_\mathrm{III} (p) \le \chi_\mathrm{III}^\mathrm{exp} \right\}.
\label{S_III}
\end{equation}
Parallel to the cases related to set II, we  consider two subsets:
\begin{equation}
S_{\mathrm{III}1} = \left\{
 p \,| \, p\, \in \, S_\mathrm{III} \,\, : \,\, m[f_0(1710)]=1690\pm 20 \,{\rm MeV}\right\}.
\label{S_III1}
\end{equation}
Selected sample points in subset (\ref{S_III1}) are given in
Table \ref{T_App_A_SIII1} in Appendix \ref{app1}.
The ideal mixing limit (\ref{ideal_cond})  leads to subset 2:
\begin{equation}
S_{\mathrm{III}2} = \left\{
 p \,| \, p\, \in \, S_\mathrm{III} \,\,: \,\,
 \left(K_{11}^{-1}\right)^2  >  0.50 \,\, \&\,\,
 \left(K_{22}^{-1}\right)^2  >  0.50 \right\}.
\label{S_III2}
\end{equation}

The details of Monte Carlo simulations II and III are similar to the details presented in section IV and we
skip them here (detailed numerical values of the physical quantities obtained in these two numerical simulations are respectively given in Tables \ref{T_App_B2} and \ref{T_App_B3} of Appendix B).   A comparison of the results for the substructure of the isosinglet states is given in Fig.~\ref{F_global_compare}:  The left figure compares the three global simulations;  the middle figure compares the results for the three subsets I1, II1 and III1 [obtained by imposing constraint on the mass of $f_0(1710)$]; and the right figure shows a comparison of the results obtained for the three subsets I2, II2 and III2 (obtained in the limit of ideal mixing for the light scalar meson nonet).   We see that the results are qualitatively stable with some moderate variations.    The $f_0(500)$, $f_0(980)$ and $f_0(1370)$ have low glue components, the first two have a considerable mixing of two- and four-quark components while $f_0(1370)$ is dominantly quark-antiquark.    The  $f_0(1500)$ and $f_0(1710)$ have substantial glue component, but the percentage of their glue is rather sensitive to the way it is probed.

To make a more quantitative comparison, we use  Eq.~(\ref{E_chi_def}) and compute its double-averaged over all data points in a set and over all target experimental quantities:
\begin{equation}
{\bar{\bar\chi}} = {1\over {N_p N_q^\mathrm{exp}}}\sum_{j=1}^{N_p} \sum_{i=1}^{N_q^\mathrm{exp}}
              \left|
              {
                { {\hat q}_i^\mathrm{exp} - q_i^\mathrm{theo}\left(p_j\right)}
                               \over
                          {{\hat q}_i^\mathrm{exp}}
              }
               \right|,
\label{E_chiave_def}
\end{equation}
where $p_j$ is a  point in a given set/subset, $N_p$ and $N_q$ are the number of points in a set/subset and the number of target quantities, respectively.    We then compare this with the average of experimental target quantities for each case computed using the definition (\ref{E_chiexp_def})
\begin{equation}
{\bar \chi}^\mathrm{exp} = { {\chi^\mathrm{exp}}\over {N_q^\mathrm{exp}}}.
\label{E_chiave_exp}
\end{equation}
The results are given in Table \ref{T_ave_compare}.   The closest (on average) that the model can get to the central values of the target experimental inputs is in simulation III, followed by simulations I and II.    The only difference between simulations I and II is that the two partial decay widths of $f_0(1370)$ are excluded in I and included in II (also included in III).  The difference between simulations II and III is that the WA102 data is included in II and excluded in III.   Since model agrees less with experiment in simulation II, we conclude that the two partial decay widths of $f_0(1370)$ \cite{Bugg:2007ja}  and WA102 data \cite{WA102} are not quite compatible, and that the present model agrees slightly better with the data in \cite{Bugg:2007ja}.

\begin{table}[h]
	\begin{center}
		\begin{tabular}{|c|c|c|c|c|}
			\hline
			Simulation & ${\bar \chi}^\mathrm{exp}$    & ${\bar{\bar\chi}}$ (global set) & ${\bar{\bar\chi}}$	 (subset 1)  & ${\bar{\bar\chi}}$ (subset 2)    \\ \hline
			I    &         35              &        19       &      17          &        25          \\ \hline
			II   &         36              &        29       &      25          &        29          \\ \hline
			 III  &         35              &        17       &      15          &        16          \\ \hline
			
		\end{tabular}
	\end{center}
	\caption{A quantitative comparison of different numerical simulations.   The averaged percent experimental uncertainties [first column; computed from Eq.~(\ref{E_chiave_exp})] is compared   with the double-averaged percent deviation of the model predictions from central experimental values [second, third and fourth columns; computed from Eq.~(\ref{E_chiave_def})]. The comparison is made over the three global sets $S_\mathrm{I}$, $S_\mathrm{II}$ and $S_\mathrm{III}$ [defined in (\ref{S_I}), (\ref{S_II}) and (\ref{S_III})]; the three subsets  $S_{\mathrm{I}1}$, $S_{\mathrm{II}1}$ and $S_{\mathrm{III}1}$ [defined in (\ref{S_I1}), (\ref{S_II1}) and (\ref{S_III1})]; and the three subsets $S_{\mathrm{I}2}$, $S_{\mathrm{II}2}$ and $S_{\mathrm{III}2}$  [defined in (\ref{S_I2}), (\ref{S_II2}) and (\ref{S_III2})].
}
\label{T_ave_compare}
\end{table}

% % % % % % % % % % % % % % % % % % % % figure 11
%
\begin{figure}[h]
	\begin{center}
%		\vskip .75cm
		%------------
\epsfxsize = 17cm
		\epsfbox{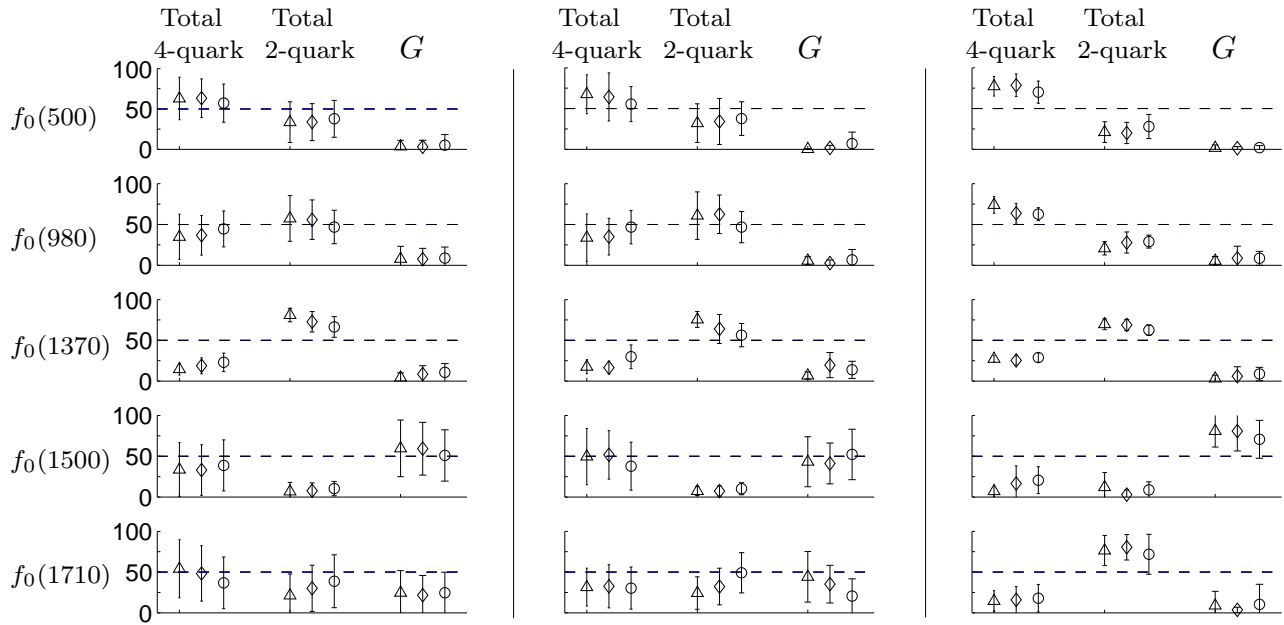}
		\caption{
Comparison of the predictions for the total four-quark, total quark-antiquark and glue of the isosinglet scalars obtained over: the three global sets $S_\mathrm{I}$, $S_\mathrm{II}$ and $S_\mathrm{III}$ [defined in (\ref{S_I}), (\ref{S_II}) and (\ref{S_III}); left];   the three subsets $S_{\mathrm{I}1}$, $S_{\mathrm{II}1}$ and $S_{\mathrm{III}1}$ [defined in (\ref{S_I1}), (\ref{S_II1}) and (\ref{S_III1}); middle]; and the three subsets $S_{\mathrm{I}2}$, $S_{\mathrm{II}2}$ and $S_{\mathrm{III}2}$ [defined in (\ref{S_I2}), (\ref{S_II2}) and (\ref{S_III2}); right].  Triangles, diamonds and circles are respectively related to simulations I, II and III.    The error bars show standard deviation around the average.     Overall the results are not very sensitive to the variation of the experimental inputs. For convenience, the 50\% lines are also shown (dashed lines).}
\label{F_global_compare}
	\end{center}
\end{figure}

\section{The scalar glueball}

The scalar field ``$G$,'' introduced in section II, was identified with a scalar glueball.   In this section we give more discussion on this field and retrospectively justify its identification with the scalar glueball.  We give more details on its phenomenological properties probed in the simulations of previous sections.

In the present framework,  the connections to the ``microscopic'' world of quarks and gluons are made indirectly via the transformation properties, mixing patterns, mass spectra  and decay properties. Since our approach  is motivated by the global description and family relations among scalar mesons,  it is natural for us to introduce the ``matter fields'' in terms of two scalar nonets (a four-quark scalar nonet $N$ and a two-quark scalar nonet $N'$) so that we can then study the family interactions between these two nonets and  have  global relations among model parameters, such as, for example, among various scalar-pseudoscalar-pseudoscalar coupling constants  that get expressed in terms of the coupling constants that describe the interaction of the two nonets in $I=1/2$ and 1 channels, as well as in the more complex case of $I=0$ channel (see Appendix B of \cite{04_F1}).
Therefore,  the scalar matter fields, i.e.~the substructures that are made out of quarks [components 1 to 4 of ${\bf F}_0$ in relation (\ref{SNS_def})] are all   contained within these two scalar nonets and their properties in the Lagrangian are dictated by the flavor SU(3) transformation and  its breakdown.    One of the advantages of these nonet templates is that they ``confine'' the  matter fields within the $N$ and $N'$.     Consequently, once the Lagrangian is formulated with these global templates for matter fields (and assuming that there are no other composite matter fields of substructures more complex than quark-antiquark and four-quarks, such as six-quarks, eight-quarks, $\ldots$)  any other SU(3)-singlet scalar field floating in the model cannot be identified with the matter fields for it leads to over-counting.     As such,  scalar field $G$  must be identified with an ``other-than-quark-composite''  field  and one possibility is to identify it with a pure  glueball field.  Since we do not have direct access into its internal substructure, we  use the external characteristics of field $G$ to see if it fits the profile of an scalar glueball.
First and foremost,  isolating the matter fields into nonets $N$ and $N'$ all scalar states below 2 GeV are accounted for.  Then if $G$ is to represent an scalar glueball, it is necessary that it couples to these matter fields as an SU(3) singlet.  Of course, this, by itself,  is not sufficient. Second,   when the model is confronted with experiment and the model parameters are explored (the results obtained in previous sections),   and consequently various bare as well as physical quantities are computed, we  see that there are  clear indications that further support the identification of $G$  with an scalar glueball. These are:

(a) In most  investigations of $f_0(500)$ and $f_0(980)$ in the literature,   it is found that they do not contain a considerable glue component.   In our investigation too, we found that $f_0(500)$ and $f_0(980)$ [and $f_0(1370)$] have   low $G$ components [see Fig.~\ref{F_raw_comps}].

(b)	In simulations of the previous sections,  the mass of the scalar field $G$  is typically distributed over the range  1.3--1.9 GeV, overlapping with the range of mass for the lightest scalar glueball found in lattice QCD.    Fig.~\ref{F_gluemass} shows the histograms for mass of field $G$ in the three global sets.  Also, in the three subsets 1, and in the three subsets 2 the masses are almost in the same range.    Fig. ~\ref{F_Glueball_mass} compares $m_G$ computed over the three global sets, the three subsets 1,  and the three subsets 2 (the averages related to simulations I, II and III are shown with triangles, diamonds and circles, respectively, and the standard deviations are shown by error bars).    We see that most averages are between the mass of $f_0(1500)$ and $f_0(1710)$.   Also shown is the intersection of all simulations (shaded band) in the range of 1.55--1.61 GeV, which is the most overlapped   range for the scalar glueball mass in the present order of this framework (in Sec. VII,  we will estimate the theoretical uncertainty due to neglecting the next set of corrections, and determine our overall estimate for $m_G$).
% % % % % % % % % % % % % % % % % % % % figure 15
%
\begin{figure}[h]
	\begin{center}
%		\vskip .75cm
		%------------
		\epsfxsize = 15cm
		\epsfbox{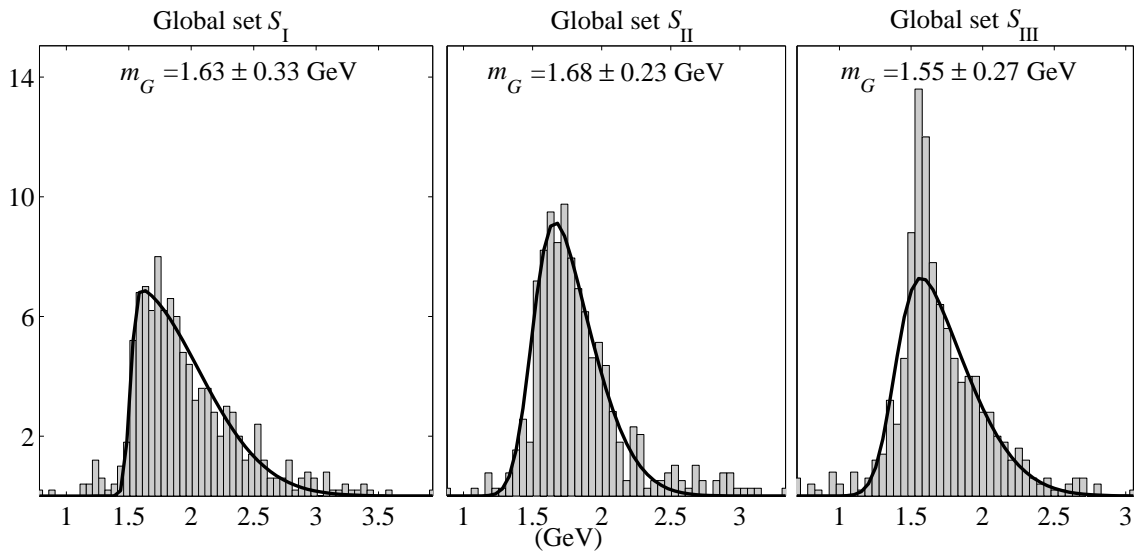}
		%\hskip 1cm
		\caption{Historgrams for $m_G$ over the three global sets $S_\mathrm{I}$, $S_\mathrm{II}$ and  $S_\mathrm{III}$ [see definitions (\ref{S_I}), (\ref{S_II}) and (\ref{S_III})].  The superimposed curves are skew Gaussian fits and the displayed numbers give the location of the peaks together with the square-root of the variances.}
\label{F_gluemass}
	\end{center}
\end{figure}

\begin{figure}[h]
	\begin{center}
		\epsfxsize = 10 cm
		\epsfbox{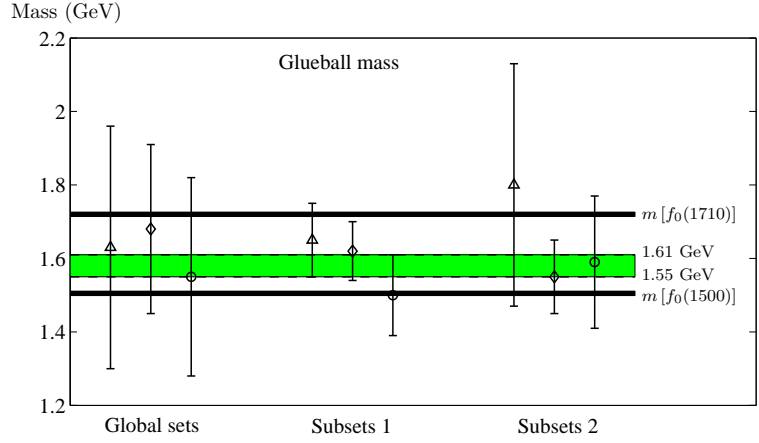}
    	\caption{
The results of numerical simulations for $m_G$ (scalar glueball mass) over the three global sets $S_\mathrm{I}$, $S_\mathrm{II}$ and $S_\mathrm{III}$ [defined in (\ref{S_I}), (\ref{S_II}) and (\ref{S_III}); left];   the three subsets $S_{\mathrm{I}1}$, $S_{\mathrm{II}1}$ and $S_{\mathrm{III}1}$ [defined in (\ref{S_I1}), (\ref{S_II1}) and (\ref{S_III1}); middle]; and the three subsets $S_{\mathrm{I}2}$, $S_{\mathrm{II}2}$ and $S_{\mathrm{III}2}$ [defined in (\ref{S_I2}), (\ref{S_II2}) and (\ref{S_III2}); right].  Triangles, diamonds and circles are respectively related to simulations I, II and III.    The error bars show standard deviation around the average.   The shaded band shows the intersection of all estimates that results in  the overall prediction of the scalar  glueball mass in the range 1.55--1.61 GeV.    The experimental mass of the $f_0(1500)$ and $f_0(1710)$ are also given \cite{pdg} and clearly entrap the scalar glueball mass.
    }
\label{F_Glueball_mass}
		\end{center}
\end{figure}

(c) The present effective Lagrangian framework,  formulated in terms of meson fields,   does not directly probe   the coupling of glueball(s) to quarks.    However, since in this approach the effective coupling of $G$ to pseudoscalar-pseudoscalar channels are computed, it becomes indirectly possible to probe the coupling of $G$ to quarks.  	The couplings of  $G$ to channels  $\pi\pi$, $K{\bar K}$, $\eta\eta$, $\eta \eta'$ and $\eta' \eta'$ in our simulations  are given in Fig.~\ref{F_Gpp}.  We see that the glueball coupling to $K{\bar K}$ channel is stronger than to $\pi\pi$ channel which seems consistent with the ``chiral suppression'' investigated in \cite{Chanowitz}.
Moreover, it is seen that  the coupling of $G$ to channels involving $\eta'$ shows considerable enhancement.    Since glue dynamics plays an essential role in understanding the physics of $\eta'$,    the larger coupling of $G$ to channels involving $\eta'$ is consistent with the  identification of scalar field $G$ with scalar glueball.

 % % % % % % % % % % % % % % % % % % % figure 14
%
\begin{figure}[h]
	\begin{center}
%		\vskip .75cm
		\epsfxsize = 17cm
		\epsfbox{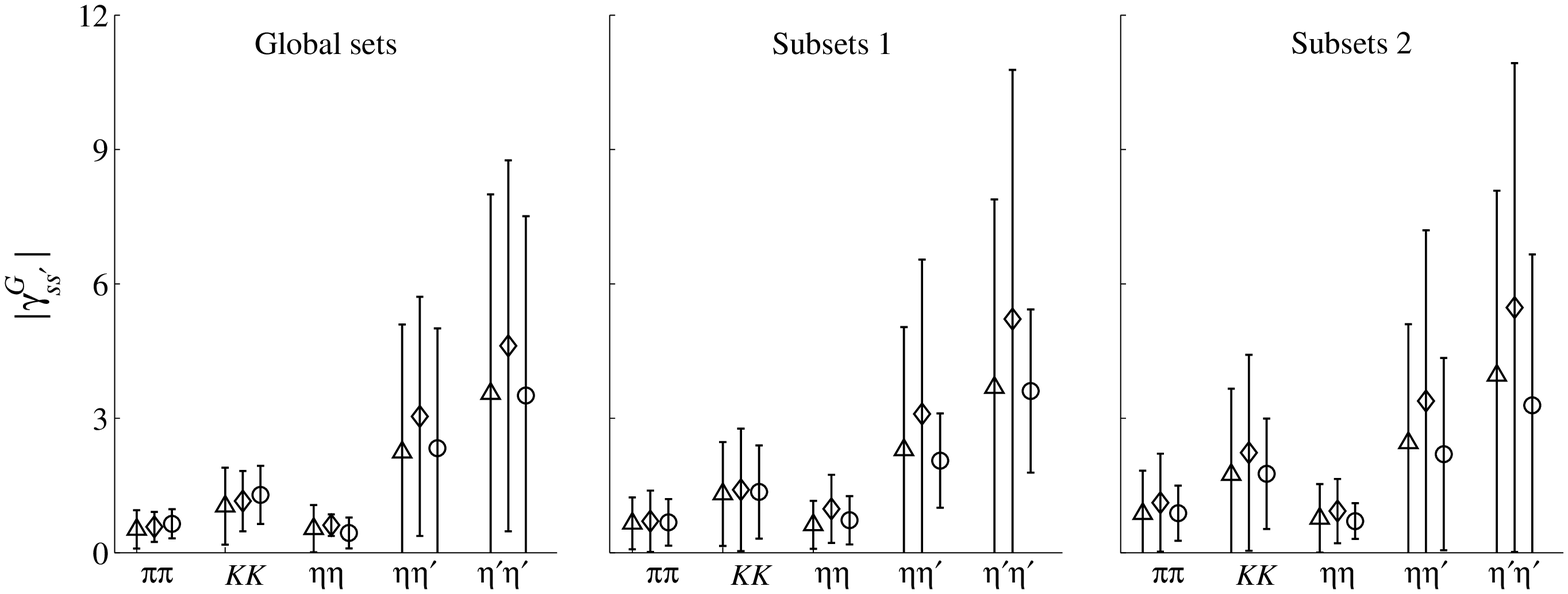}
    	\caption{
    The magnitude of the coupling of $G$ to pseudoscalar-pseudoscalar channels ($\left|\gamma^G_{ss'}\right|$ in GeV$^{-1}$)
over the three global sets $S_\mathrm{I}$, $S_\mathrm{II}$ and $S_\mathrm{III}$ [defined in (\ref{S_I}), (\ref{S_II}) and (\ref{S_III}); left].    Triangles, diamonds and circles are respectively related to simulations I, II and III.    The error bars show standard deviation around the average.   Field $G$ couples stronger to $K{\bar K}$  than to $\pi\pi$ and  its coupling to channels including $\eta'$ increases significantly.
}
\label{F_Gpp}
		\end{center}
\end{figure}

(d) A the present leading order of this framework,  the glueball mixing with quarkonia  is SU(3) symmetric  and, in principle, can only distinguish the  two-quark from the four-quark states but not  the difference between $u$, $d$ and $s$.   In this limit,   the number of simulations that result in a larger coupling of $G$ to the quark-antiquark nonet $N'$ have an edge over those that give a larger coupling of $G$ to the four-quark nonet $N$. Fig.~\ref{F_f_over_e} shows the percentages of simulations for which $|f/e|>1$ [see Eq.~(\ref{L_mix_I0})].     For example, we see that in global sets I, II and III, respectively 45\%, 52\% and 59\% of simulations have $|f/e| > 1$. Since the glueball mass typically  spread over the range of 1.3--1.9, one would  expect that the glueball couples stronger to quark-antiquark nonet $N'$ than to the four-quark nonet $N$.

\begin{figure}[h]
	\begin{center}
		\epsfxsize = 10 cm
		\epsfbox{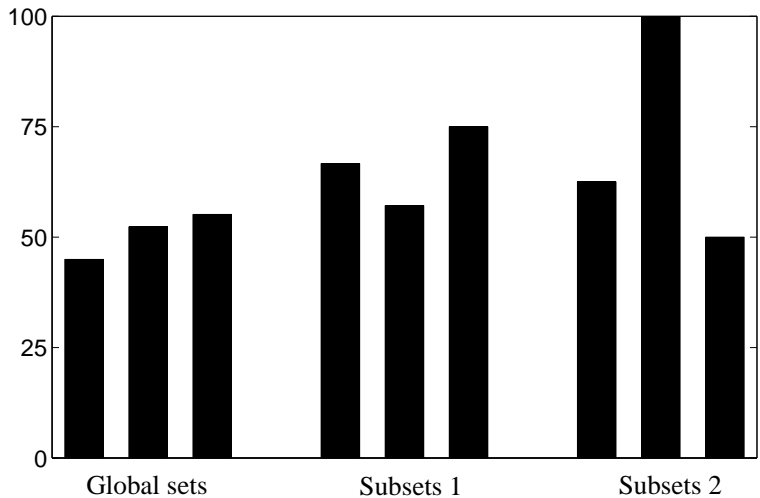}
    	\caption{Comparing the glueball coupling to quark-antiquark nonet $N'$ over its coupling to four-quark nonet $N$.    Histograms show the percentages of simulations that resulted in $|f/e|>1$ [see Eq.~(\ref{L_mix_I0})] for the three global sets $S_\mathrm{I}$, $S_\mathrm{II}$ and $S_\mathrm{III}$ [defined in (\ref{S_I}), (\ref{S_II}) and (\ref{S_III}); left];   the three subsets $S_{\mathrm{I}1}$, $S_{\mathrm{II}1}$ and $S_{\mathrm{III}1}$ [defined in (\ref{S_I1}), (\ref{S_II1}) and (\ref{S_III1}); middle]; and the three subsets $S_{\mathrm{I}2}$, $S_{\mathrm{II}2}$ and $S_{\mathrm{III}2}$ [defined in (\ref{S_I2}), (\ref{S_II2}) and (\ref{S_III2}); right].
    Overall the simulations favor $|f/e|>1$.    }
\label{F_f_over_e}
		\end{center}
\end{figure}

In summary,   although the present framework is not meant to directly connect to the microscopic world of quarks and gluons,  and therefore it cannot be proved (microscopically) that $G$ is a scalar glueball,   nevertheless, phenomenological arguments presented here,  clearly agree with its identification as a scalar glueball.

%%%%%%%%%%%%%%%%%%%%%%%%%%%%%%%%%%%%%%%%%%%%%%%%

\section{Summary and discussion}

%%%%%%%%%%%%%%%%%%%%%%%%%%%%%%%%%%%%%%%%%%%%%%%%

The issue of the substructure of $f_0(1500)$ and $f_0(1710)$, particularly their glue content, was the main focus of the present work.   We showed, both within the three global sets [(\ref{S_I}), (\ref{S_II}) and (\ref{S_III})] as well as within the three subsets 1 [(\ref{S_I1}), (\ref{S_II1}) and (\ref{S_III1})] and the three subsets 2
[(\ref{S_I2}), (\ref{S_II2}) and (\ref{S_III2})], that glue is almost entirely absorbed by $f_0(1500)$ and $f_0(1710)$ (overall, more than 75\% of an effective scalar glueball field was absorbed by these two states). However,  identifying the percentages of glue in either state can be  an excruciatingly difficult task, mainly due to unestablished experimental data on properties of  some of the isosinglet scalars, many with large uncertainties (and at times conflicting).  The  strategy of this work in extracting the glue contents (within the existing experimental uncertainties) has been to establish a global framework for all isosinglets, isodoublets and isotriplet scalars below 2 GeV and derive underlying mixing patterns and family relations as a mean to probe the contents of all scalar states in general, and the glue component of $f_0(1500)$ and $f_0(1710)$, in particular.   This strategy is of course a ``double-edge sword:'' On the one hand,  the entanglement of all scalars together (not only the isosinglets but also the isodoublets and isotriplets)  results in added complications and new stumbling blocks that would otherwise be absent in an isolated study.  On the other hand, an important benefit is that the family  relations developed in this process make it possible to  impose  some of the known (or widely believed) features of some of the scalars to extract information on others.

\begin{figure}[h]
	\begin{center}
		\vskip .75cm
		\epsfxsize = 12 cm
		\epsfbox{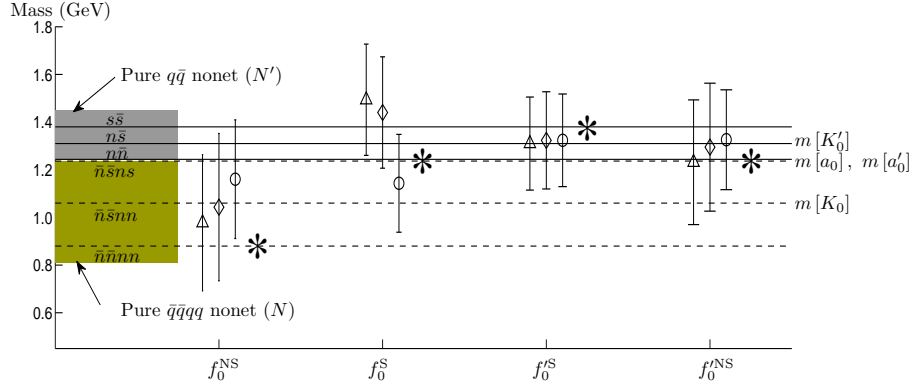}
		\caption{
The bare  masses of the members of the four-quark scalar meson nonet $N$ and the quark-antiquark scalar meson nonet $N'$ in the present model. The mass of the four isosinglet scalars $f_0^{\rm NS}$, $f_0^{\rm S}$, ${f'}_0^{\rm S}$ and ${f'}_0^{\rm NS}$ are computed
over the three subsets $S_\mathrm{I1}$, $S_\mathrm{II1}$ and $S_\mathrm{III1}$ [defined in (\ref{S_I1}), (\ref{S_II1}) and (\ref{S_III1});    triangles, diamonds and circles respectively represent the averages in simulations I, II and III and the error bars represent the standard deviations around the averages].   The stars represent the isosinglet scalar masses computed from the relations (\ref{Bare_masses_Np}) and (\ref{Bare_masses_N}).  The masses of the $I=1/2,\ 1$ members of the two nonets are also shown.}
\label{F_Bare_masses}
	\end{center}
\end{figure}

Specifically, We applied a nonlinear chiral Lagrangian model that includes a quark-antiquark and a four-quark nonet of scalar mesosns and an scalar glueball.    This model has already been applied to various low-energy scattering and decays and has resulted in a coherent picture for the low-energy data \cite{HSS1,BFSS1,BFSS2,Mec}.   Using an iterative Monte Carlo method,  we probed the 14 free parameters of the model by fitting to more than 20 experimental  inputs on isosinglet scalars below 2 GeV together with constrains from isodoublets and isotriplet properties.   We determined three global  sets of points $S_\mathrm{I}$, $S_\mathrm{II}$ and $S_\mathrm{III}$ [Eqs.~(\ref{S_I}), (\ref{S_II}) and (\ref{S_III})] in the 14 dimensional parameter space that gave an overall agreement with experimental data on all isosinglet states.        In probing the glue component of the $f_0(1500)$ and $f_0(1710)$,  after determining the global sets,  we first computed the quark and glue components of all isosinglet  states below 2 GeV,  and then studied family relations among them.    Without any additional assumptions,  the average components over the three global sets  showed that $f_0(1500)$ and $f_0(1710)$ are almost exclusively the main glue holders.     Even though in most of our numerical simulations $f_0(1500)$ dominated the glue contents, the accuracy of our model (at its present order) does not allow us to confidently favor $f_0(1500)$ versus $f_0(1710)$ (a discussion of the accuracy of the prsent order of the model, together with an order of magnitude estimate of  the size of next set of corrections are given below).

We then further imposed additional constraint on the mass of $f_0(1710)$ on the three global sets to control the resulting physical masses.    This resulted in the three subsets 1 [i.e.~$S_{\mathrm{I}1}$, $S_{\mathrm{II}1}$ and $S_{\mathrm{III}1}$;  Eqs.~(\ref{S_I1}), (\ref{S_II1}) and (\ref{S_III1}), respectively] which we used to reevaluate the quark and glue components, but qualitatively the results remained consistent with those obtained with the three global sets.   The glueball mass was also estimated in this limit in the range 1.55--1.61 GeV.

We exploited the underlying correlations among isosinglet states as a mean to further probe the substructure of $f_0(1500)$ and $f_0(1710)$.   Imposing the anticipated  characteristics of  $f_0(500)$ and $f_0(980)$ to not contain large glue contents (which is generally observed in the literature to be the case), favors $f_0(1500)$ as the state with a major glue content. We also further examined  a limit in which the lightest isosinglets $f_0(500)$ and $f_0(980)$ get close to their ideally four-quark mixing limit.   This led to   the three subsets 2 [i.e.~$S_{\mathrm{I}2}$, $S_{\mathrm{II}2}$ and $S_{\mathrm{III}2}$;  Eqs.~(\ref{S_I2}), (\ref{S_II2}) and (\ref{S_III2}), respectively]. In this limit, again  $f_0(1500)$ was favored to contain the dominant glue component [which was approximately estimated above 70\%].

We end by giving an estimate of the theoretical uncertainty due to neglecting (for simplicity) the higher order corrections.      The foundation of this analysis has been the mixings of a bare (unmixed) light four-quark nonet $N$ with a heavier bare quark-antiquark nonet $N'$ and a scalar glueball $G$.   The Lagrangian discussed in section II is developed based on chiral symmetry and its breakdown and the transformation properties of the effective meson fields are understood from the basic transformation properties of the schematic quark fields.     The schematic quark configurations, imply certain mass ordering expected for each bare nonet.    In the quark-antiquark scalar nonet, we expect:
\begin{equation}
m[{f'}^{\rm NS}_0] \approx m[a'_0] < m[K'_0] < m[{f'}^{\rm S}_0]
\end{equation}
where ${f'}^{\rm NS}_0$, $a_0'$, $K_0'$, and ${f'}^{\rm S}_0$ belong to  $N'$ and have quark substructures
\begin{eqnarray}
{f'}^{\rm NS}_0 &:& n{\bar n} \nonumber \\
a_0'  &:& n{\bar n} \nonumber \\
{K_0'} &:& n{\bar s} \nonumber \\
{f'}^{\rm S}_0 &:& s{\bar s} \nonumber \\
\label{Np_config}
\end{eqnarray}
Moreover, assuming schematic quarks (which is the basis of the underlying flavor symmetry of the Lagrangian) and solely based on counting the strange and non-strange quarks, we (roughly) expect:
\begin{eqnarray}
m[{f'}^{\rm NS}_0] &\approx& m[a'_0] \nonumber \\
m[K'_0] - m[a'_0] &\approx& m[{f'}^{\rm S}_0] - m[K'_0]
\label{Bare_masses_Np}
\end{eqnarray}
which only measures the number of strange quarks in each state.

\begin{table}[t]
	\begin{center}
		\begin{tabular}{|c|c|}
			\hline
			Our overall  estimate	            & $ 1580\pm 180 $\\ \hline \hline
			G. S. Bali \textit{et al.} \cite{L1}	& $ 1550 \pm 50 $\\ \hline
			H. Chen \textit{et al.} \cite{L2}    & $ 1740 \pm 71 $ \\ \hline
			C. J. Morningstar and M. J. Peardon \cite{Morningstar_99}& $ 1730 \pm 50 \pm 80 $ \\ \hline
			A. Vaccarino and D. Weingarten \cite{Vaccarino:1999ku}& $ 1648 \pm 58 $ \\ \hline
			F. E. Close and Q. Zhao  \cite{L5}   & $ 1464 \pm 47$ and $ 1519 \pm 41$ \\ \hline
			M. Loan \textit{et al.} \cite{L6}    & $ 1654 \pm 83 $ \\ \hline
			Y. Chen \textit{et al.} \cite{L7}    & $ 1710 \pm 50 \pm 80 $ \\ \hline
			E. Gregory \textit{et al.} \cite{Gregory:2012hu}  & $ 1795 \pm 60 $ \\ \hline
		\end{tabular}
		\caption{Comparison of the  glueball mass estimated in the  present framework  with several predictions of lattice QCD.}
		\label{T_mG_lattice}
	\end{center}
\end{table}

Similarly, we expect a mass ordering among the members of four-quark nonet $N$ (which is inverted compared to that of quark-antiquark nonet $N'$):
\begin{equation}
m[f^{\rm NS}_0] < m[K_0] < m[a_0]  \approx m[f^{\rm S}_0]
\end{equation}
where $f^{\rm NS}_0$, $a_0$, $K_0$ and $f^{\rm S}_0$ belong to $N$ and have quark substructures
\begin{eqnarray}
f^{\rm NS}_0 &:& {\bar n} {\bar n} n n \nonumber \\
{K_0} &:& {\bar n} {\bar s} n n  \nonumber \\
a_0  &:&  {\bar n} {\bar s} n s  \nonumber \\
f^{\rm S}_0 &:&  {\bar n} {\bar s} n s
\label{N_config}
\end{eqnarray}
Again assuming an ``schematic spectroscopy'' and counting the strange and non-strange quarks, we (roughly) expect:
\begin{eqnarray}
m[K_0] - m[f^{\rm NS}_0] &\approx&  m[a_0] - m[K_0] \nonumber \\
 m[a_0] &\approx&  m[f^{\rm S}_0]
\label{Bare_masses_N}
\end{eqnarray}
In the study of the $I=1/2,\ 1$ within this mixing framework \cite{Mec},  the bare masses of $m[K_0]$, $m[a_0]$, $m[K'_0]$ and $m[a'_0]$ are determined [see Eq.~(\ref{a0k0_m_bare})].     These are also plotted in Fig.~\ref{F_Bare_masses} together with the $I=0$ bare masses which are computed in two different ways:  (i) from numerical simulation of this analysis (triangles, diamonds and circles) and (ii)  from
relations (\ref{Bare_masses_Np}) and (\ref{Bare_masses_N}) (stars)  expected from the corresponding schematic quark configurations (\ref{Np_config}) and (\ref{N_config}).
Since our simulations did not target the bare masses and instead targeted  the physical masses, decay widths and decay ratios, we interpret the difference between these two sets of calculation of $I=0$ bare masses a measure of the theoretical uncertainty  of our work (due to neglecting higher order terms in this Lagrangian).
Using (\ref{Bare_masses_Np}) and (\ref{Bare_masses_N}), the four  masses shown by stars in Fig.~\ref{F_Bare_masses} are (in GeV):
\begin{equation}
m^*_1 = 0.88 \,\, , \,\,  m^*_2 = 1.24 \,\, , \,\,
m^*_3 = 1.38 \,\, , \,\, m^*_4 = 1.24.
\end{equation}
The average theoretical uncertainty (averaged over the four stared masses as well as the three sets of simulations) can be estimated
\begin{equation}
{\bar \Delta} = {1\over {12}} \sum_{j} \sum_{i=1}^4 { {\left| {\bar m}_j[{\bf F}_{0,i}] - m^*_i\right|}\over m^*_i}
\approx 11\%
\end{equation}
where ${\bf F}_0$ is defined in (\ref{SNS_def}), and ${\bar m}_j$ represents the averaged mass over the three subsets 1 [i.e.~$S_{\mathrm{I}1}$, $S_{\mathrm{II}1}$, $S_{\mathrm{III}1}$ defined in Eqs.~(\ref{S_I1}),  (\ref{S_II1}) and  (\ref{S_III1}), respectively].

In the leading order of the model we found that the favored range of glueball mass is 1.55--1.61 GeV.   Therefore,  including the 11\% theoretical uncertainty, our overall estimate for the glueball mass is $1.58 \pm 0.18$ GeV and is compared with several lattice QCD estimates in Table \ref{T_mG_lattice}.   We expect that the higher order corrections such as the higher order SU(3) symmetry breaking terms will reduce this uncertainty.  These corrections have been formally identified and their effects on the mass matrices of $I$=1/2,1,0 have been worked out in \cite{04_F1},  however,  their determination from fits to available data will require a considerable extension of the parameter space of the model, and the leading order results presented here pave the way for investigating the larger parameter space.   We intend to pursue these extensions in future works.

\section*{Acknowledgments}
\vskip -.5cm
A.H.F.~wishes to thank the Physics Dept.~of Shiraz University for its hospitality in Summer of 2012 where this work was initiated.

\clearpage

\appendix

\section{Sample points in set $S_\mathrm{I}$ \label{app1}}

In this appendix we have given sample points in the three subsets $S_{\mathrm{I}1}$, $S_{\mathrm{II}1}$ and $S_{\mathrm{III}1}$ [defined in (\ref{S_I1}), (\ref{S_II1}) and (\ref{S_III1})].   These subsets result in  the isosinglet physical masses that are closer to their target values.   It is interesting to note that although the Lagrangian parameters spread over a range of correlated values (even changing sign),  the physical parameter  $m_G$ given in the three tables (similar to the physical parameters given in the next appendix)  varies over a physical range.

\begin{table}[h]
   	\begin{center}
   		\begin{tabular}{|c||r|r|r|r|r|r|r|r|r|r|}
    			\hline
  $c$&$ 0.3370  $&$  0.3952  $&$  0.0145   $&$ 0.3622  $&$  0.0243  $ \\ \hline
  $d$&$ -0.0150 $&$  -0.0254 $&$   0.0146  $&$ -0.0251 $&$   0.0201 $\\ \hline
  $c^\prime$&$ -0.1497 $&$  -0.1216 $&$   0.2715  $&$  0.1313 $&$  -0.1673 $ \\ \hline
  $d^\prime$&$  0.0187 $&$   0.0160 $&$  -0.0240  $&$ -0.0236 $&$   0.0081 $ \\ \hline
  $m_G$&$  1.6998 $&$   1.7407 $&$   1.4901  $&$  1.7345 $&$   1.6064 $\\ \hline
  $\rho$&$ -0.2097 $&$  -0.0441 $&$   0.0603  $&$ -0.2507 $&$  -0.5877 $ \\ \hline
  $e$&$ -0.1074 $&$  -0.0460 $&$  -0.2558  $&$  0.0343 $&$  -0.2076 $ \\ \hline
  $f$&$  0.1214 $&$   0.2593 $&$   0.1499  $&$  0.3861 $&$   0.0062 $ \\ \hline \hline
  $B$&$  0.5185 $&$  -0.0342 $&$  -0.2162  $&$  0.2433 $&$   0.2655 $ \\ \hline
  $D$&$ -0.0955 $&$   0.2531 $&$  -0.6200  $&$ -0.6983 $&$  -0.6260 $ \\ \hline
  $B^\prime$&$ -1.6051 $&$  -1.5578 $&$  -0.2952  $&$ -0.4635 $&$  -1.9995 $ \\ \hline
  $D^\prime$&$ -1.6865 $&$  -1.6900 $&$   5.9564  $&$ -0.7670 $&$   0.7092 $ \\ \hline
  $E$&$ -0.5866 $&$  -0.1665 $&$  -0.7889  $&$  0.2909 $&$  -0.3853 $ \\ \hline
  $F$&$ -0.3537 $&$   2.2655 $&$   1.0031  $&$ -4.3858 $&$   0.3655 $ \\ \hline \hline
  $\chi$&$  3.0710 $&$   3.9327 $&$   3.9756  $&$  4.1966 $&$   4.5207 $ \\ \hline
          \end{tabular}
     \end{center}
\caption{Sample points in subset $S_\mathrm{I1}$ [defined in (\ref{S_I1})],  $\chi_\mathrm{I}^\mathrm{exp}=7.3$}
\label{T_App_A_SI1}
\end{table}

\begin{table}[h]
	\begin{center}
		\begin{tabular}{|c||r|r|r|r|r|r|r|r|r|r|}
			\hline
$c$ & $ 0.2609  $ & $  0.1456 $ & $  -0.2117  $ & $  0.1022  $ & $  0.2636  $  \\ \hline
$d$ & $-0.0003  $ & $  0.0075 $ & $   0.0516  $ & $  0.0155  $ & $ -0.0178  $   \\ \hline
$c^\prime$ & $ 0.2433  $ & $  0.0660 $ & $   0.0025  $ & $ -0.2672  $ & $  0.1270  $   \\ \hline
$d^\prime$ & $-0.0231  $ & $ -0.0186 $ & $   0.0184  $ & $  0.0126  $ & $ -0.0185  $   \\ \hline
$m_G$ & $ 1.7445  $ & $  1.5946 $ & $   1.7237  $ & $  1.5704  $ & $  1.5855  $   \\ \hline
$\rho$ & $-0.2718  $ & $ -0.1367 $ & $  -0.3336  $ & $ -0.0016  $ & $ -0.4982  $   \\ \hline
$e$ & $ 0.0492  $ & $ -0.0488 $ & $   0.0546  $ & $  0.1009  $ & $  0.1733  $  \\ \hline
$f$ & $-0.2221  $ & $ -0.5522 $ & $  -0.0988  $ & $ -0.0395  $ & $ -0.1686  $  \\ \hline \hline
$B$ & $ 0.0875  $ & $  0.0538 $ & $   0.6674  $ & $  0.5891  $ & $  0.3232  $   \\ \hline
$D$ & $-1.9439  $ & $ -1.8003 $ & $  -1.3296  $ & $ -1.1793  $ & $ -0.7720  $   \\ \hline
$B^\prime$ & $-0.7580  $ & $  1.4509 $ & $  -1.3466  $ & $ -3.0306  $ & $ -0.5641  $  \\ \hline
$D^\prime$ & $-0.8105  $ & $ -1.8948 $ & $   0.3781  $ & $ -1.3564  $ & $  2.4016  $  \\ \hline
$E$ & $-0.2582  $ & $  1.4825 $ & $  -0.1311  $ & $ -0.5332  $ & $ -0.1580  $   \\ \hline
$F$ & $-4.6696  $ & $  2.9563 $ & $  -0.3990  $ & $  0.1563  $ & $ -3.8550  $   \\ \hline \hline
$\chi$ & $ 5.9322  $ & $  6.0866 $ & $   6.2091  $ & $  6.7777  $ & $  7.2839  $  \\ \hline
      \end{tabular}
    \end{center}
    \caption{Sample points in subset $S_\mathrm{II1}$ [defined in (\ref{S_II1})], $\chi_\mathrm{II}^\mathrm{exp}=8.2$}
    \label{T_App_A_SII1}
\end{table}

\begin{table}[h]
	\begin{center}
		\begin{tabular}{|c||r|r|r|r|r|r|r|r|r|r|}
			\hline
$c$ & $-0.1312 $ & $  -0.1896 $ & $  -0.1695  $ & $  0.0465  $ & $  0.0968 $ \\ \hline
$d$ & $ 0.0258 $ & $   0.0208 $ & $   0.0385  $ & $  0.0233  $ & $ -0.0077 $ \\ \hline
$c^\prime$ & $-0.1023 $ & $   0.2979 $ & $   0.0637  $ & $ -0.0504  $ & $  0.1417 $ \\ \hline
$d^\prime$ & $ 0.0121 $ & $  -0.0281 $ & $  -0.0186  $ & $  0.0160  $ & $ -0.0122 $ \\ \hline
$m_G$ & $ 1.4885 $ & $   1.3729 $ & $   1.5075  $ & $  1.6338  $ & $  1.5627 $ \\ \hline
$\rho$ & $-0.4765 $ & $   0.0389 $ & $   0.1461  $ & $ -0.3293  $ & $ -0.5419 $ \\ \hline
$e$ & $ 0.3334 $ & $  -0.4492 $ & $   0.1004  $ & $  0.0662  $ & $  0.1091 $ \\ \hline
$f$ & $ 0.4258 $ & $   0.2080 $ & $  -0.1142  $ & $  0.0527  $ & $  0.1577 $ \\ \hline \hline
$B$ & $-1.0804 $ & $  -1.4822 $ & $   0.4289  $ & $  0.4434  $ & $  0.3742 $ \\ \hline
$D$ & $-3.2734 $ & $  -1.5545 $ & $  -2.9775  $ & $ -0.8613  $ & $  1.4400 $ \\ \hline
$B^\prime$ & $-2.7836 $ & $   0.2785 $ & $  -0.3578  $ & $ -1.4455  $ & $ -0.5599 $ \\ \hline
$D^\prime$ & $-2.0592 $ & $  -2.2582 $ & $  -2.5229  $ & $ -0.2597  $ & $ -0.8833 $ \\ \hline
$E$ & $ 1.6391 $ & $  -2.5225 $ & $   0.2925  $ & $  0.1295  $ & $ -0.4453 $ \\ \hline
$F$ & $ 2.4595 $ & $  -1.6122 $ & $  -5.1189  $ & $  0.2258  $ & $ -1.1873 $ \\ \hline \hline
$\chi$ & $ 1.9395 $ & $   3.9535 $ & $   4.1088  $ & $  4.1678  $ & $  5.3078 $ \\ \hline
      \end{tabular}
    \end{center}
    \caption{Sample points in subset $S_\mathrm{III1}$ [defined in (\ref{S_III1})], $\chi_\mathrm{III}^\mathrm{exp}=5.6$}
    \label{T_App_A_SIII1}
\end{table}
%*******
\section{Physical quantities \label{app2}}

In this appendix we have given the numerical values for the physical quantities computed over the three global sets
$S_\mathrm{I}$, $S_{\mathrm{II}}$ and $S_{\mathrm{III}}$ [defined in (\ref{S_I}), (\ref{S_II}) and (\ref{S_III})],
the three subsets $S_{\mathrm{I}1}$, $S_{\mathrm{II}1}$ and $S_{\mathrm{III}1}$ [defined in (\ref{S_I1}), (\ref{S_II1}) and (\ref{S_III1})], and the three subsets $S_{\mathrm{I}2}$, $S_{\mathrm{II}2}$ and $S_{\mathrm{III}2}$ [defined in (\ref{S_I2}), (\ref{S_II2}) and (\ref{S_III2})].

\begin{table}[t]
	\begin{center}
		\begin{tabular}{|c|c|c|c|c|}
			\hline
			Short notation & Target value& Global set I &  Subset I1 &  Subset I2 \\ \hline
			$m_1$ &  $ 400-550 $ \cite{pdg}& $521 \pm 114$ & $598 \pm 154  $ & $ 634\pm167 $ \\ \hline
			$m_2$ & $  990 \pm 20 $ \cite{pdg}& $966 \pm 160$ & $ 972\pm 90 $   & $ 1011\pm146 $  \\ \hline
			$m_3$ & $  1312 $ \cite{WA102} & $1500 \pm 92$ &  $ 1478\pm 53 $  & $ 1442\pm33 $ \\ \hline
			$m_4$ &  $ 1502 $ \cite{WA102}& $1804 \pm 258$ & $ 1638\pm100 $  & $ 1761\pm264 $   \\ \hline
			$m_5$&  $ 1727 $ \cite{WA102}& $2828 \pm 633$ &  $ 1773\pm30 $  &  $ 2834\pm729$  \\ \hline \hline
			$\Gamma^3_{\frac{\pi\pi}{KK}}$  &  $ 2.17 \pm    0.9 $ \cite{WA102}& $ 2.67 \pm 1.39 $ & $ 2.30\pm0.66 $  &$ 4.48\pm2.80 $    \\ \hline
			$\Gamma^3_{\frac{\eta\eta}{KK}}$ &  $ 0.35 \pm    0.3 $ \cite{WA102}& $ 0.11 ^{+ 0.14}_{-0.11} $ & $ 0.12 ^{+0.18}_{-0.12} $  & $ 0.09 ^{+0.15}_{-0.09} $   \\ \hline
			$\Gamma^4_{\frac{\pi\pi}{\eta\eta}}$ &  $ 5.56 \pm    0.93$ \cite{WA102}& $ 6.00 \pm 1.14$& $5.56\pm1.07  $  &$ 6.59\pm0.88 $   \\ \hline
			$\Gamma^4_{\frac{KK}{\pi\pi}}$ & $  0.33 \pm    0.07 $ \cite{WA102}& $ 0.31 \pm 0.07 $& $ 0.34\pm0.04 $  & $ 0.34\pm0.03 $  \\ \hline
			$\Gamma^4_{\frac{\eta \eta'}{\eta \eta}}$ &  $ 0.53 \pm   0.23 $ \cite{WA102}&$ 0.42 \pm 0.12$& $0.36\pm0.08  $  &$ 0.50\pm0.09 $  \\ \hline
			$\Gamma^5_{\frac{\pi\pi}{KK}}$&  $ 0.20 \pm    0.03 $ \cite{WA102}&$ 0.23 \pm 0.04$&  $ 0.19\pm0.02 $  & $ 0.20\pm0.02 $  \\ \hline
			$\Gamma^5_{\frac{\eta \eta}{KK}}$&  $ 0.48 \pm    0.19 $ \cite{WA102}&$ 0.48 \pm 0.06$& $ 0.41\pm0.14 $  &$ 0.49\pm0.01 $  \\ \hline \hline
			$\Gamma^1_{\pi\pi}$&    $ 400-700 $ \cite{pdg} &$16 ^{+21}_{-16}$& $ 38^{+41}_{-38} $  & $ 43 ^{+45}_{-43} $   \\ \hline
			${\widetilde\Gamma}^1_{\pi\pi}$&   N/A &$482\pm257$& $ 340\pm293 $  & $ 288 ^{+308}_{-288} $   \\ \hline
			$\Gamma^2_{\pi\pi}$&  $ 40-100 $ \cite{pdg} &$65 \pm 23$&  $ 68\pm11 $  & $ 82\pm30 $  \\ \hline
			$\Gamma^3_{\pi\pi}$& N/A &$243 \pm 54$& $ 214\pm45 $  &$ 222\pm92 $     \\ \hline
			$\Gamma^3_{KK}$& N/A &$109 \pm 47$& $98\pm28  $  & $ 73\pm56 $   \\ \hline
			$\Gamma^4_{\pi\pi}$&  $ 38 \pm    5 $ \cite{pdg} &$31\pm 4$& $32\pm6  $   & $32\pm4  $ \\ \hline
			$\Gamma^4_{KK}$&  $ 9 \pm    2 $ \cite{pdg}&$10\pm 1 $&  $  11\pm3$  & $11\pm2  $ \\ \hline
			$\Gamma^4_{\eta\eta}$&  $ 6 \pm    1 $ \cite{pdg}&$5 \pm 1$& $ 6\pm2 $  & $ 5\pm1 $   \\ \hline
			$\Gamma^4_{\eta\eta\prime}$ &  $ 2.1 \pm    1.0  $ \cite{pdg}&$2.1 \pm 0.5$& $ 2.1\pm 0.5 $  &$ 2.5\pm 0.5 $   \\ \hline
			$\Gamma^5_{\pi\pi}$&  $ 26 ^{+29}_{-26} $ \cite{oller}&$21 \pm 4$&  $ 20\pm3 $  &  $ 18\pm1 $ \\ \hline
			$\Gamma^5_{KK}$&  $ 80    \pm 40 $ \cite{oller}&$94 \pm 15$& $ 105\pm18 $   &  $ 86\pm8 $ \\ \hline
			$\Gamma^5_{\eta\eta}$ &  $ 48 \pm 35  $ \cite{oller}&$45 \pm 7$&  $ 42\pm11 $  & $ 43\pm4 $   \\ \hline
		\end{tabular}
	\end{center}
\caption{Numerical results for the masses, decay ratios and decay widths computed in simulation I.    The target values are given in the second column together with the averages and standard deviations computed over: the global set $S_\mathrm{I}$ [defined in (\ref{S_I})] (third column);   the subset  $S_{\mathrm{I}1}$ [defined in (\ref{S_I1})] (fourth column); and subset $S_{\mathrm{I}2}$ [defined in (\ref{S_I2})] (fifth column).   For the decay width of $f_0(500)$ to two pions, both the bare decay width $\left(\Gamma^1_{\pi\pi}\right)$ as well as the physical decay width $\left({\widetilde\Gamma}^1_{\pi\pi}\right)$ in which the effects of the final-state interaction of pions are taken into account, are given.  In the second column, ``N/A'' denotes the quantities that have not been targeted in the simulation, nevertheless, their values are computed as by products.   The short notations in the first column are defined in Table \ref{quantities1}.   For comparison convenience,  the last seven target values are presented in the form ``central value $\pm$ uncertainty'' format (their original reported values are given in Table \ref{quantities1}).}
\label{T_App_B1}
\end{table}
%%%%%%%%%
\begin{table}[t]
	\begin{center}
		\begin{tabular}{|c|c|c|c|c|}
			\hline
			Short notation & Target value& Global set II &  Subset II1 &  Subset II2 \\ \hline
			$m_1$&   400--550  \cite{pdg}& $ 605  \pm  138 $  &   $ 551 \pm   135 $   &  $  623  \pm   174 $ \\ \hline
			$m_2$& $  990 \pm 20 $ \cite{pdg}& $ 1003  \pm  187 $   &  $ 1047  \pm  196 $  &   $ 972   \pm    62$ \\ \hline
			$m_3$ & $  1312 $ \cite{WA102}& $ 1502  \pm  75 $   &  $ 1492   \pm 98 $  &  $  1436   \pm 23$  \\ \hline
			$m_4$ &  $ 1502 $ \cite{WA102}& $ 1729  \pm  176 $   &  $ 1616   \pm 91 $   &  $ 1580  \pm  62 $  \\ \hline
			$m_5$ &  $ 1727 $ \cite{WA102}& $ 2620  \pm  525 $   &  $ 1772   \pm 45  $  &  $ 2328   \pm 402 $ \\ \hline \hline
			$\Gamma^3_{\frac{\pi\pi}{KK}}$ &  $ 2.17 \pm    0.9 $ \cite{WA102}& $ 1.75  \pm   0.76 $   &  $ 1.90  \pm    1.27  $&   $  2.11  \pm    0.49  $\\ \hline
			$\Gamma^3_{\frac{\eta\eta}{KK}}$&  $ 0.35 \pm    0.3 $ \cite{WA102}& $ 0.13  ^ {+0.14}_{-0.13}  $  &  $ 0.18  \pm   0.17 $  &  $ 0.02    \pm 0.02 $ \\ \hline
			$\Gamma^4_{\frac{\pi\pi}{\eta\eta}}$ &  $ 5.56 \pm    0.93$ \cite{WA102}& $ 6.05  \pm   1.06  $  & $ 5.84   \pm  0.85 $   & $ 5.86   \pm  0.56 $   \\ \hline
			$\Gamma^4_{\frac{KK}{\pi\pi}}$& $  0.33 \pm    0.07 $ \cite{WA102} & $ 0.32  \pm   0.16 $   & $ 0.29   \pm  0.04 $   &  $ 0.32  \pm   0.02 $\\ \hline
			$\Gamma^4_{\frac{\eta \eta'}{\eta \eta}}$&  $ 0.53 \pm   0.23 $ \cite{WA102}&$ 0.41   \pm   0.11 $  & $  0.39  \pm    0.10  $&   $ 0.44  \pm   0.06 $ \\ \hline
			$\Gamma^5_{\frac{\pi\pi}{KK}}$&  $ 0.20 \pm    0.03 $ \cite{WA102}&$ 0.23   \pm  0.04 $   & $ 0.21   \pm  0.02  $  &  $ 0.22   \pm   0.03 $   \\ \hline
			$\Gamma^5_{\frac{\eta \eta}{KK}}$ &  $ 0.48 \pm    0.19 $ \cite{WA102}&$ 0.48   \pm  0.05  $  & $  0.51   \pm   0.05  $&   $  0.47 \pm    0.02  $ \\ \hline \hline
			$\Gamma^1_{\pi\pi}$&    400--700  \cite{pdg}& $ 38  ^{+ 50}_{-38} $  &   $ 48   ^{+77}_{-48} $  &   $ 54^{+57}_{-54} $ \\ \hline
			${\widetilde \Gamma}^1_{\pi\pi}$&    N/A & $308 \pm 283$& $ 420 \pm 294 $  & $ 306^{+335}_{-306} $   \\ \hline			
			$\Gamma^2_{\pi\pi}$ &   40--100  \cite{pdg} & $ 91  \pm   46 $  &   $ 116  \pm   79  $&    $ 137 \pm    73 $ \\ \hline
			$\Gamma^3_{\pi\pi}$&  $ 59 \pm 25  $ \cite{Bugg:2007ja}&$  170 \pm    45 $ &   $  153 \pm    55 $&     $ 171\pm     29 $   \\ \hline
			$\Gamma^3_{KK}$& $  80 \pm   35 $ \cite{Bugg:2007ja}& $ 113 \pm    45  $&    $ 100  \pm   43  $&    $ 83  \pm  13 $ \\ \hline
			$\Gamma^4_{\pi\pi}$&  $ 38 \pm    5 $ \cite{pdg}& $ 32  \pm   4  $  &  $ 35    \pm 5  $   & $ 30     \pm2 $  \\ \hline
			$\Gamma^4_{KK}$&  $ 9 \pm    2 $ \cite{pdg}&$  10  \pm   3  $  &  $ 10    \pm 2 $    & $ 9     \pm 1 $  \\ \hline
			$\Gamma^4_{\eta\eta}$&  $ 6 \pm    1 $ \cite{pdg}&$  5   \pm  1 $    & $ 6     \pm 1 $     &$ 5     \pm 1 $ \\ \hline
			$\Gamma^4_{\eta\eta\prime}$&  $ 2.1 \pm    1.0 $ \cite{pdg}& $ 2.2   \pm  0.5  $   & $ 2.3     \pm 0.4  $    &$ 2.2     \pm 0.2 $ \\ \hline
			$\Gamma^5_{\pi\pi}$&  $ 26 ^{+29}_{-26} $ \cite{oller}& $ 22  \pm   4 $   &  $ 19   \pm  2  $  &  $ 23  \pm   4 $\\ \hline
			$\Gamma^5_{KK}$&  $ 80    \pm 40 $ \cite{oller}& $ 99  \pm   15 $  &   $ 91  \pm   10 $  &   $ 101\pm     9 $\\ \hline
			$\Gamma^5_{\eta\eta}$&  $ 48 \pm 35  $ \cite{oller}& $ 47  \pm   6  $  &  $ 47   \pm  4 $   &  $ 47  \pm  3 $    \\ \hline
		\end{tabular}
	\end{center}
\caption{Numerical results for the masses, decay ratios and decay widths computed in simulation II.    The target values are given in the second column together with the averages and standard deviations computed over: the global set $S_\mathrm{II}$ [defined in (\ref{S_II})] (third column);   the subset  $S_{\mathrm{II}1}$ [defined in (\ref{S_II1})] (fourth column); and subset $S_{\mathrm{II}2}$ [defined in (\ref{S_II2})] (fifth column). For the decay width of $f_0(500)$ to two pions, both the bare decay width $\left(\Gamma^1_{\pi\pi}\right)$ as well as the physical decay width $\left(\widetilde{\Gamma}^1_{\pi\pi}\right)$ in which the effects of the final-state interaction of pions are taken into account, are given.   The short notations in the first column are defined in Table \ref{quantities1}. For comparison convenience,  the last nine target values are presented in the form ``central value $\pm$ uncertainty'' format (their original reported values are given in Table \ref{quantities1}).}
\label{T_App_B2}
\end{table}
% % % % % %
\begin{table}[t]
	\begin{center}
		\begin{tabular}{|c|c|c|c|c|}
			\hline
			Short notation& Target value & Global set III &  Subset III1 &  Subset III2 \\ \hline
			$m_1$ &   400--550  \cite{pdg}&     $ 626  \pm  133 $ &   $ 605  \pm  131 $  &  $ 686  \pm  132 $ \\ \hline
			$m_2$ & $  990 \pm 20 $ \cite{pdg}&     $ 964  \pm  166 $ &   $ 937  \pm  146 $  &  $ 1015  \pm  90 $    \\ \hline
			$m_3$ &  $ 1300 \pm 15  $ \cite{Bugg:2007ja}&     $ 1469  \pm  55 $ &   $ 1448  \pm  54 $  &  $ 1438  \pm  17 $ \\ \hline
			$m_4$ &  $ 1505 \pm 6  $ \cite{pdg}&     $ 1645  \pm  140 $ &   $ 1574  \pm  57 $  &  $ 1632  \pm  110 $  \\ \hline
			$m_5$ &  $ 1690 \pm 20  $ \cite{oller} &     $ 2296  \pm  422 $ &   $ 1717  \pm  61 $  &  $ 2157  \pm  183 $  \\ \hline \hline
			$\Gamma^3_{\frac{\pi\pi}{KK}}$  & N/A&     $ 1.51  \pm  1.27 $ &   $ 1.41  \pm  0.88 $  &  $ 1.99  \pm  1.72 $ \\ \hline
			$\Gamma^3_{\frac{\eta\eta}{KK}}$ & N/A&     $ 0.07  ^{+0.15}_{-0.07} $ &   $ 0.11  ^{+0.16}_{-0.11} $  &  $ 0.03  \pm  0.03 $ \\ \hline
			$\Gamma^4_{\frac{\pi\pi}{\eta\eta}}$ & N/A&     $ 7.52  \pm  1.74 $ &   $ 7.51  \pm  1.86 $  &  $ 8.29  \pm  1.23 $  \\ \hline
			$\Gamma^4_{\frac{KK}{\pi\pi}}$ & N/A&     $ 0.26  ^{+0.37}_{-0.26} $ &   $ 0.24  \pm  0.05 $  &  $ 0.24  \pm  0.01 $ \\ \hline
			$\Gamma^4_{\frac{\eta \eta'}{\eta \eta}}$& N/A&    $ 0.43  \pm  0.42 $ &   $ 0.38  \pm  0.07 $  &  $ 0.44  \pm  0.06 $  \\ \hline
			$\Gamma^5_{\frac{\pi\pi}{KK}}$& N/A&    $ 0.38  \pm  0.33 $ &   $ 0.31  \pm  0.12 $  &  $ 0.28  \pm  0.09 $ \\ \hline
			$\Gamma^5_{\frac{\eta \eta}{KK}}$& N/A&    $ 0.69  \pm  0.64 $ &   $ 0.55  \pm  0.24 $  &  $ 0.55  \pm  0.06 $  \\ \hline \hline
			$\Gamma^1_{\pi\pi}$&    400--700  \cite{pdg}&     $ 62  ^{+83}_{-62} $ &   $ 53  ^{+63}_{-53} $  &  $ 68  \pm  44 $  \\ \hline
			${\widetilde\Gamma}^1_{\pi\pi}$&    N/A &$262 ^{+275}_{-262}$& $ 303\pm270 $  & $ 153 ^{+260}_{-153} $   \\ \hline			
			$\Gamma^2_{\pi\pi}$&   40--100  \cite{pdg}&     $ 81  \pm  35 $ &   $ 71  \pm  30 $  &  $ 84  \pm  30 $   \\ \hline
			$\Gamma^3_{\pi\pi}$&  $ 59 \pm 25  $ \cite{Bugg:2007ja}&     $ 136  \pm  51 $ &   $ 117  \pm  51 $  &  $ 133  \pm  46 $   \\ \hline
			$\Gamma^3_{KK}$& $  80 \pm   35 $ \cite{Bugg:2007ja}&     $ 121  \pm  53 $ &   $ 100  \pm  41 $  &  $ 97  \pm  46 $ \\ \hline
			$\Gamma^4_{\pi\pi}$&  $ 38 \pm    5 $ \cite{pdg}&     $ 39  \pm  4 $ &   $ 40 \pm  8 $  &  $ 38  \pm  1 $  \\ \hline
			$\Gamma^4_{KK}$&  $ 9 \pm    2 $ \cite{pdg}&     $ 9  \pm  1 $ &   $ 9  \pm  1 $  &  $ 9  \pm  1 $  \\ \hline
			$\Gamma^4_{\eta\eta}$&  $ 6 \pm    1 $ \cite{pdg}&     $ 5  \pm  1 $ &   $ 5  \pm  1 $  &  $ 5  \pm  1 $  \\ \hline
			$\Gamma^4_{\eta\eta\prime}$&  $ 2.1 \pm    1.0 $ \cite{pdg}&     $ 2.2  ^{+2.7}_{-2.2} $ &   $ 2.1  \pm  0.2 $  &  $ 2.1  \pm  0.1 $  \\ \hline
			$\Gamma^5_{\pi\pi}$&  $ 26 ^{+29}_{-26} $ \cite{oller}&     $ 26  \pm  3 $ &   $ 27  \pm  2 $  &  $ 24  \pm  8 $ \\ \hline
			$\Gamma^5_{KK}$ &  $ 80    \pm 40 $ \cite{oller}&     $ 80  \pm  29 $ &   $ 103  \pm  65 $  &  $ 85  \pm  7 $\\ \hline
			$\Gamma^5_{\eta\eta}$&  $ 48 \pm 35  $ \cite{oller}&     $ 48 \pm  6 $ &   $ 47  \pm  8 $  &  $ 47  \pm  4 $ \\ \hline
		\end{tabular}
	\end{center}
\caption{Numerical results for the masses, decay ratios and decay widths computed in simulation III.    The target values are given in the second column together with the averages and standard deviations computed over: the global set $S_\mathrm{III}$ [defined in (\ref{S_III})] (third column);   the subset  $S_{\mathrm{III}1}$ [defined in (\ref{S_III1})] (fourth column); and subset $S_{\mathrm{III}2}$ [defined in (\ref{S_III2})] (fifth column).   For the decay width of $f_0(500)$ to two pions, both the bare decay width $\left(\Gamma^1_{\pi\pi}\right)$ as well as the physical decay width $\left({\widetilde \Gamma}^1_{\pi\pi}\right)$ in which the effects of the final-state interaction of pions are taken into account, are given.In the second column, ``N/A'' denotes the quantities that have not been targeted in the simulation, nevertheless, their values are computed as by products.   The short notations in the first column are defined in Table \ref{quantities2}. For comparison convenience,  the last nine target values are presented in the form ``central value $\pm$ uncertainty'' format (their original reported values are given in Table \ref{quantities2}).}
\label{T_App_B3}
\end{table}

\end{document}